\documentclass{PoS}

\title{Time lag in transient galactic and extragalactic accreting sources}

\ShortTitle{Time lag in transient accreting sources}
\author{\speaker{Franco Giovannelli}%\thanks{}
\\
        INAF - Istituto di Astrofisica e Planetologia Spaziali, Via del Fosso del Cavaliere, 100, 00133 Roma, Italy\\
        E-mail: \email{franco.giovannelli@iaps.inaf.it}}
\author{Gennady S. Bisnovatyi-Kogan\\IKI, Moscow, Russian Federation\\
E-mail: \email{gkogan@iki.rssi.ru}}
\abstract{X-ray binaries are cauldrons of fundamental physical processes which appear along practically the whole electromagnetic spectrum.
The sub-class of X-ray transient sources show multifrequency behaviour which deserve particular attention in order to understand the causing physics.
These binary systems consist of a compact star and an optical star, therefore there is a mutual
influence between these two stars that drive the low energy (LE) (i.e. radio, IR, optical) and high energy (HE) (i.e. UV, X-ray, $\gamma$-ray) processes. The LE processes are produced mostly on the optical star and the HE processes mostly on the compact star, typically a neutron star. Thus it appears evident that through the study of LE processes it is possible to understand also the HE processes and vice versa.
In this paper we will discuss this problem starting from the experimental evidence of a delay between LE and HE processes detected for the first time in the X-ray/Be system A0535+26/HDE245770 (e.g. Giovannelli \& Sabau-Graziati, 2011; Giovannelli, Bisnovatyi-Kogan \& Klepnev, 2013 (here after GBK13); Giovannelli et al., 2015b). This delay is common in cataclysmic variables (CVs) and other binary systems with either a neutron star or a black hole.

Since a delay between LE processes and HE processes has been experimentally observed in several active galactic nuclei (AGNs), we will discuss also the tidal disruption of stars by massive BHs, following the original idea of Rees (1998): stars in galactic nuclei can be captured or tidally disrupted by a central black hole. Some debris would be ejected at high speed, the remainder would be swallowed by the hole, causing a bright flare lasting at most a few years.

The outline of this paper is:
\begin{itemize}
  \item Antecedent fact
  \item Introduction
  \item X-ray Binary Systems
  \item Old \& News from the transient X-ray/Be system A0535+26/HDE245770
  \item The model for Galactic Accreting Sources
  \item The model for AGNs
  \item Discussion \& Conclusions
\end{itemize}
}

\FullConference{Accretion Processes in Cosmic Sources - II - APCS2018\\
		3-8 September 2018\\
		Saint Petersburg, Russian Federation}

\begin{document}

\section{Antecedent Fact}

Figure 1 clearly explain all the mysteries of our Universe (Giovannelli, 2000). People who are able to read this sentence can understand that "{\bf The truth is written in the book of the Nature.
We must learn to read this book}".

%%%%%%%%%%%%%%%%%% Fig. 1 %%%%%%%%%%%%%%%%%%%
\begin{figure}%[!ht]%[!hbp]%[!ht]%[!hbp]%[!ht]%[!hbp]%[h][h!]
\centering
\includegraphics[width=0.5\textwidth]{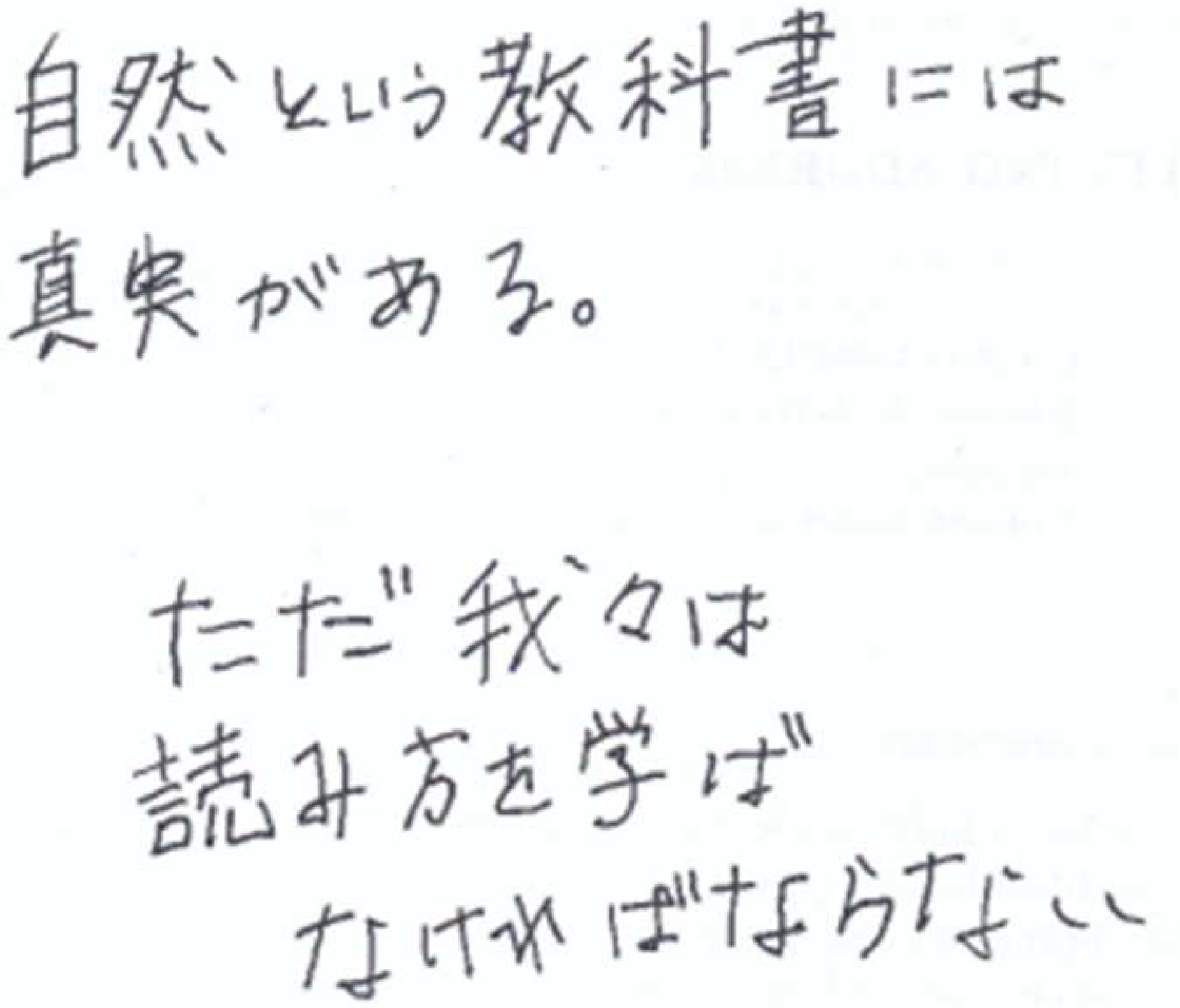}
\caption{Understanding our Universe (adopted from Giovannelli, 2000).}
\label{fig1}
\end{figure}
%%%%%%%%%%%%%%%%%%%%%%%%%%%%%%%%%%%%%%%%%%%%%

The experiments provide the basic alphabet, immersed in an apparently chaotic soup, but necessary to understand the nature. From that soup we must extract words and phrases to compose the book of the nature. In other words, the data coming from the experiments constitute the basic alphabet that we use for constructing models that attempt to describe the nature. But we have a lot of models for interpreting the experimental data by the light of science. Depending on the hypotheses the results could run against the experiments. Then, in order to be acceptable, models can take into account and justify {\bf ALL the available data}.

The same concept was expressed in much more incisive terms by Richard Phillips Feynman -- Nobel laureate in Physics in 1965 -- also known as {\it The Great Explainer}: {\bf It doesn't matter how beautiful your theory is, it doesn't matter how smart you are. If it doesn't agree with experiment, it's wrong}.

The talk originating this paper had not been scheduled in the original Final Program of the workshop because the detailed description of the models that we will present is published in the proceedings of the Saint Petersburg workshop 2016 by G.S. Bisnovatyi-Kogan and F. Giovannelli, and other their publications. This talk came from the hole created in the program due to the sudden absence of Nazar Ikhsanov and Mateusz Wisniewicz. Therefore, instead of the details of the models, we will present the genesis of this work to the benefit above all of the young participants.

This paper is a demonstration of how we tried to read the book of the nature.

Undoubtedly the advent of new generation experiments ground-- and space--based have given a strong impulse for verifying current theories, and for providing new experimental inputs for developing a new physics for going, probably, over the standard model (SM). Recent results coming from Active Physics Experiments (APEs) -- experiments in which we try to reproduce in a laboratory the physical conditions of processes occurring in nature we want to understand -- and Passive Physics Experiments (PPEs) -- experiments by which we observe the nature -- have opened such a new path.

An extensive review on the situation about the knowledge of the physics of our Universe has been recently published by Giovannelli \& Sabau-Graziati (2016). The reader interested is invited to look at that paper.

The correct procedure in passive physics experiments is:
\begin{itemize}
  \item to observe and collect experimental data;
  \item to analyze these data without any a priori bias;
  \item to attempt their interpretation on the base of current models;
  \item if not possible, it is mandatory to search for other models that cannot be "ad hoc".
\end{itemize}

This seems trivial, but, unfortunately, it is not so.

\section{Introduction}

Important observations started early in
the past century with the discovery of
inexplicable effects - the supernovae.
They had been observed in ancient times
but it was only with the establishing  of
the stellar and galactic distance scales
that their true enormity was realized,
namely the release of $\geq 10^{50}$ erg within
a matter of days.

The beginning of the Space Astrophysics Era is commonly located around the end of the fifties of the last century with the first space experiments, in the energy range 0.2--0.5 MeV, on board balloons. They were devoted to the detection of $\gamma$-rays generated in solar activity (Peterson \& Winckler, 1958). But actually $\gamma$-ray astronomy was born in the last year of the XIXth century with the discoveries of penetrating gamma radiation (Villard, 1900), and the atmospheric ionization (Wilson, 1900). Wilson suggested that the extraterrestrial gamma radiation could be responsible for the atmospheric ionization. With balloon flights Hess (1912) demonstrated the extraterrestrial and extra-solar origin of the ionizing radiation, which was called {\it Cosmic Rays}.

Until 1927 it was thought that cosmic rays consisted of $\gamma$-rays. Thanks to the discovery of the dependence of the cosmic ray flux on the geomagnetic latitude during a trip from Java to Genoa made by Clay (1927), it was recognized
that the composition of cosmic rays included charged particles. Later Hayakawa (1952) determined the contribution of $\gamma$-rays to the composition of cosmic rays as less than 1\%.
The experiments outside the atmosphere started in 1946, soon after the end of the second world war, when the Naval Research Laboratory (NRL) launched a V2 rocket with a payload which observed the Sun's UV spectrum.

Since that time many space experiments were prepared and several fundamental results were reached. In our opinion the actual beginning of the Space Era for studying the Universe is the year 1962. An X-ray experiment -- prepared by Giacconi, Gursky, Paolini \& Rossi -- launched on board an Aerobee rocket discovered a strong X-ray emission
from an extra-solar object, namely Sco X-1 (Giacconi et al., 1962). After this first historical experiment many others were launched on board rockets and later balloons and satellites. These experiments lead to our knowledge of an X-ray sky hitherto unknown which started to give experimental proofs of the first theories of Baade \& Zwicky (1934) about the possible existence of neutron stars.

Indeed,
Baade \& Zwicky (1934)  first suggested that
the supernova was the result of the transition
from a normal star to a neutron star.
The essential point (Zwicky, 1939)  being that
the energy releases in such a process is
comparable to the change in gravitational
potential energy of a star, which collapses
from its "normal" size of $\sim 10^6$ Km down to
the size of a neutron star of $\sim$ 10 Km.

In the 1950's, Burbidge et al. (1957) with
their works on stellar nucleosynthesis
suggested realistic models of stars prior to
supernova explosion.
The supernova process was seen as the result
of catastrophic change of state occurring in
the core of a highly evolved star, e.g. the
transformation of an iron core into a helium
core.

 Contrary, Cameron (1958) suggested that
this degenerate iron core would collapse to
a neutron core through inverse beta decay.

The discovery (by chance) of the first X-ray source
(Sco X-1) (Giacconi et al., 1962)\footnote{This was the first measurement that originated  the Nobel Price of Riccardo Giacconi.}
accelerated the
studies on neutron  stars, until that
Zel'dovich \& Guseinov (1965) suggested the
presence of an unseen massive companion
in a binary system.

Space orbiting observatories with larger and more sophisticated experiments -- from Uhuru launched in 1970 (Giacconi et al., 1971) up to HEAO-1 launched in 1977 (Wood et al., 1984) -- discovered the most luminous galactic and extragalactic X-ray sources, such as pulsars, X-ray binaries, supernova remnants (SNRs), bursters, and active galactic nuclei (AGNs). But the qualitative jump in the observational capabilities was obtained with the HEAO-2 satellite (Einstein) (Giacconi et al., 1979) in which the X-ray focussing optics of the instruments enhanced the sensitivity in the soft X-ray range by a factor of about 1000 with respect to the old generation of detectors. Also the angular resolution was improved up to $\sim$ 2 arcsec.

This allowed a re-definition of the positions of the already known X-ray sources and the discovery
of a large number of weaker ones, such as the normal galaxies and normal stars spread on the entire HR diagram (Vaiana et al., 1981).
The detected X-ray fluxes from these stars are definitively larger than those expected from theories of formation and heating of stellar coronae.

This led to a revolution in the comprehension of the role of the star rotation and of the magnetic
field in the turbulent transport of energy from the nucleus to the external parts of a star.

Giovannelli (2016) published an extensive review about X-ray binary systems.  Briefly we can summarize the main characteristics of these systems.

\section{X-ray binary systems}

The trivial definition of X-ray binaries (XRBs) is that they are binary systems emitting X-rays. However it has been largely demonstrated that X-ray binary systems emit energy in IR, Optical, UV, X-ray, Gamma-ray and sometimes they show also valuable radio emission. They can be divided in different sub-classes
\begin{itemize}
  \item {\bf High Mass X-ray Binaries (HMXB)} in which the optical companion is an early type giant or supergiant star and the collapsed object is a neutron star or a black hole. They are concentrated around the galactic plane.
The mass transfer is usually occurring via stellar wind; they show hard pulsed X-ray emission (from 0.069 to 1413 s) with KT $\geq 9$ keV; typical X-ray luminosity is ranging from $10^{34}$ to $10^{39}$ erg s$^{-1}$, and the ratio of X-ray to optical luminosity is $\sim 10^{-3}$--10. The HMXBs can be divided in two sub-classes
\begin{itemize}
  \item Hard Transient X-ray Sources (HXTS) in which the neutron star is eccentrically (e $\sim 0.2$--0.5) orbiting around a V-III luminosity-class Be star (P$_{\rm orb} > 10$ days); they show strong variable pulsed hard X-ray emission (L$_{\rm Xmax}$/L$_{\rm Xmin} >  100$) with KT $\geq 17$ keV, and P$_{\rm spin}$ ranging from 0.069 to 1413 s; L$_{\rm X} = 10^{34} - 10^{39}$ erg s$^{-1}$.
  \item Permanent X-ray Sources in which the neutron star or black hole is circularly orbiting (e $\sim 0$) around a giant or supergiant OB star (P$_{\rm orb} < 10$ days); they show an almost steady permanent pulsed hard X-ray emission (L$_{\rm Xmax}$/L$_{\rm Xmin} \ll  100$), and P$_{\rm spin}$ ranging from 0.069 to 1413 s; L$_{\rm X} \sim 10^{37}$ erg s$^{-1}$.
      \item Supergiant X-ray Binaries (SGXBs): obscured sources, which display huge amount of low energy absorption produced by the dense wind of the supergiant companion, surrounded by a weakly magnetized neutron star.
      \item Supergiant Fast X-ray transients (SFXT), a subclass of SGXBs and a new subclass of transients in which the formation of transient accretion discs could be partly responsible for the flaring activity in systems with narrow orbits. They show L$_{\rm Xpeak} \approx 10^{36}$ erg s$^{-1}$, and L$_{\rm Xquiecence} \approx 10^{32}$ erg s$^{-1}$.
\end{itemize}
%\item Anomalous X-ray Pulsars (AXPs) in which the optical counterparts probably are not OB and Be stars. They show a soft-hard X-ray emission with KT $\sim$ 0.4--4 KeV, L$_{\rm Xmax}$/L$_{\rm Xmin}\approx 10$, and
 %     L$_{\rm X}$/L$_{\rm opt} \sim$ 0.001--10, being L$_{\rm X} = 10^{34} - 10^{36}$ erg s$^{-1}$. On the contrary to the former two classes the rotational period of the pulsar is limited in a very narrow range: P$_{\rm spin} \sim$ 6--12 s.
%\end{itemize}
  \item  {\bf Low Mass X-ray Binaries (LMXB)} in which the optical companion is a low-mass-late-type star and the collapsed object is a neutron star or a black hole (P$_{\rm orb}$ from 41 min to 11.2 days). They are concentrated in the globular clusters, and in the halo around
     % around the galactic plane and especially in
       the galactic center. The mass transfer in these systems is usually occurring via Roche lobe overflow. Their emission in soft X-ray range is usually not pulsed with KT $\leq 9$ keV. Their X-ray luminosity is ranging from 10$^{36}$ to $10^{39}$ erg s$^{-1}$ and L$_{\rm X}$/L$_{\rm opt} \sim 10^2$--$10^4$; many LMXBs show Quasi Periodic Oscillations (QPOs) between 0.02 and 1000 seconds and few of them also pulsed X-ray emission, such as Her X1, 4U 1626-27 and GX 1+4.

      Many LMXB show transient behaviour in the form irregular X-ray bursts, when their luminosity increase several tens or hundreds times. During these luminous stages steady periodical signals, with milliseconds (ms) period, have been observed in several of them. In few of them ms X-ray pulsars have been discovered in quiescent stages between bursts. The ms X-ray pulsars in LMXB form a link between binary X-ray  sources and recycled binary radio pulsars, with ms periods and low magnetic fields (Bisnovatyi-Kogan and Komberg, 1974), which are formed on the place of these LMXB after ceasing of accretion, due to evolution of the companion star, transforming into low mass white dwarf, or a giant degenerate planet.
     % Unlike HMXBs, LMXBs rarely harbour an X-ray pulsar. This is because the magnetic fields of the neutron stars in %LMXBs are 10$^{-1}$ -- 10$^{-4}$ times those in HMXBs, and so the accreted material is not funneled onto the %polar caps.
  \item {\bf Cataclysmic Variables (CVs)} in which the optical companion is a low-mass-late-type star and the compact object is a white dwarf. The detected CVs are spread roughly around the solar system at distance of  200-300 pc. Orbital periods are ranging from tens of minutes to about ten hours. The mass transfer is occurring either via Roche lobe overflow or via accretion columns or in an intermediate way depending on the value of the magnetic field. Typical X-ray luminosity is ranging from $10^{32}$ to $10^{34}$ erg s$^{-1}$. Updated reviews about CVs are those by Giovannelli (2008) and Giovannelli \& Sabau-Graziati (2015a);
  \item {\bf RS Canum Venaticorum (RS CVn) type systems}, in which no compact objects are present and the two components are a F or G hotter star and a K star. Typical X-ray luminosity is ranging from $10^{30}$ to $10^{31}$ erg s$^{-1}$. Usually in the current literature they are excluded from the class of X-ray binaries since historically they were discovered as X-ray emitters only with the second generation of X-ray experiments.
\end{itemize}

Figure 2 shows a compendium of the characteristics of the X-ray binaries (adapted from Giovannelli, 2015).

%%%%%%%%%%%%%%%%%% Fig. 2 %%%%%%%%%%%%%%%%%%%
\begin{figure}%[!ht]%[!hbp]%[!ht]%[!hbp]%[!ht]%[!hbp]%[h][h!]
\centering
\includegraphics[width=0.8\textwidth]{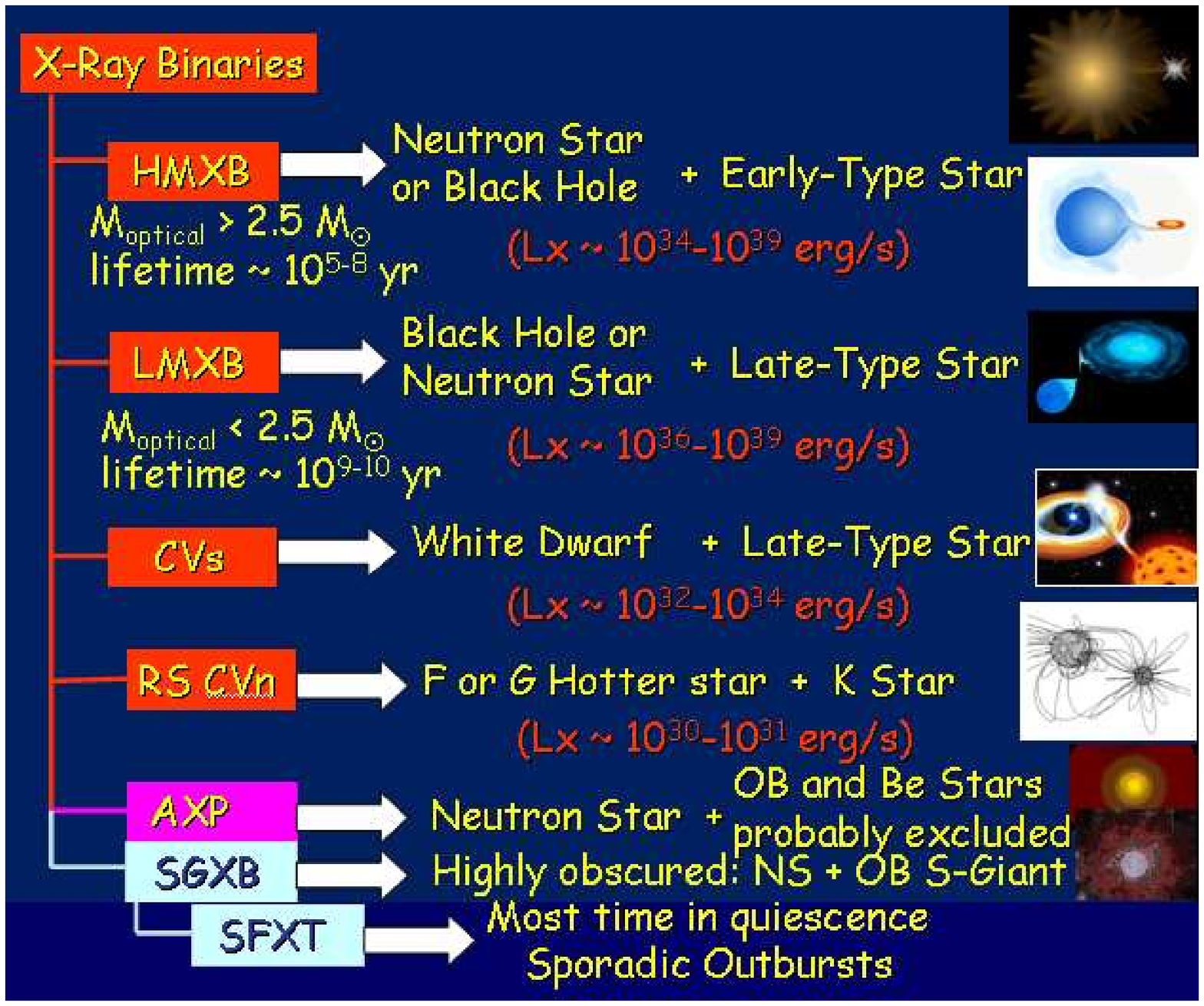}
\caption{Classification of X-ray binaries (adapted from Giovannelli, 2015).}
\label{fig1a}
\end{figure}
%%%%%%%%%%%%%%%%%%%%%%%%%%%%%%%%%%%%%%%%%%%%%

In binary systems there are essentially two ways for accreting matter from one star to the other: via accretion disk or via stellar wind (Giovannelli \& Sabau-Graziati, 2001, adapted from Blumenthal \& Tucker, 1974) (left panel of Fig 3). But in some cases there is a third way which is a mixture between the two, as for instance in eccentric binary systems close to the periastron passage where a temporary accretion disk can be formed around the neutron star (e.g. Giovannelli \& Zi\'{o}{\l}kowski (1990), like shown in the right panel of Fig. 3 (Giovannelli \& Sabau-Graziati, 2001, after Nagase, 1989).

%%%%%%%%%%%%%%%%%%% FIGURE 3 %%%%%%%%%%%%%%%%%
\begin{figure*}%[!hbp]%[!ht]%[!hbp]%[!ht]%[!hbp]%[h]
\begin{center}
\includegraphics[width=13cm]{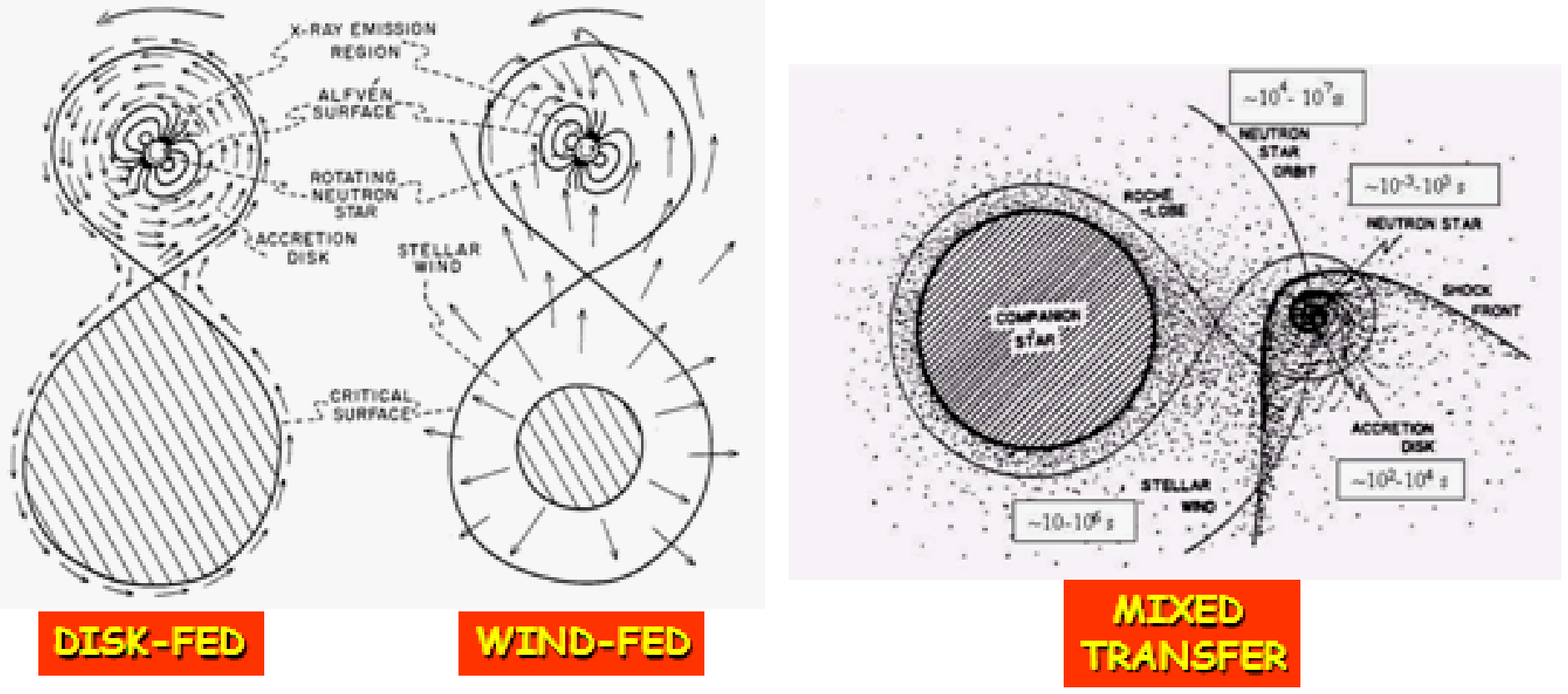}
\caption{Left panel: accretion in X-ray binary systems disk-fed and wind-fed (adopted from Giovannelli \& Sabau-Graziati, 2001, adapted from Blumenthal \& Tucker, 1974). Right panel: mixed transfer (adopted from Giovannelli \& Sabau-Graziati, 2001, after Nagase, 1989).  }
\end{center}
\end{figure*}
%%%%%%%%%%%%%%%%%%%%%%%%%%%%%%%%%%%%%%%%%%%%%%

XRBs are the best laboratory for the study of accreting processes thanks to their relative high luminosity in a large part of the electromagnetic spectrum.
For this reason, multifrequency observations are fundamental in understanding their morphology and the physics governing their behaviour.

Because of the strong interactions between the optical companion and collapsed object, low and high energy processes are strictly related.

Often, it is easier to perform observations of low energy processes (e.g. in radio, near-infrared (NIR) and optical bands) since the experiments are typically ground-based, on the contrary to observations of high energy processes, for which experiments are typically space-based.

\subsection{High Mass X-ray Binaries}

Among the X-ray binaries, the class of High Mass X-ray Binaries (HMXBs) constitutes an important group for studying the interactions either via stellar wind either via accretion disk between the optical and the compact companions.

Figure 4 shows schematically the classification of HMXBs (adopted from Giovannelli, 2015).

%%%%%%%%%%%%%%%%%%% FIGURE 4 %%%%%%%%%%%%%%%%%
\begin{figure*}%[!hbp]%[!ht]%[!hbp]%[!ht]%[!hbp]%[h]
\begin{center}
\includegraphics[width=13cm]{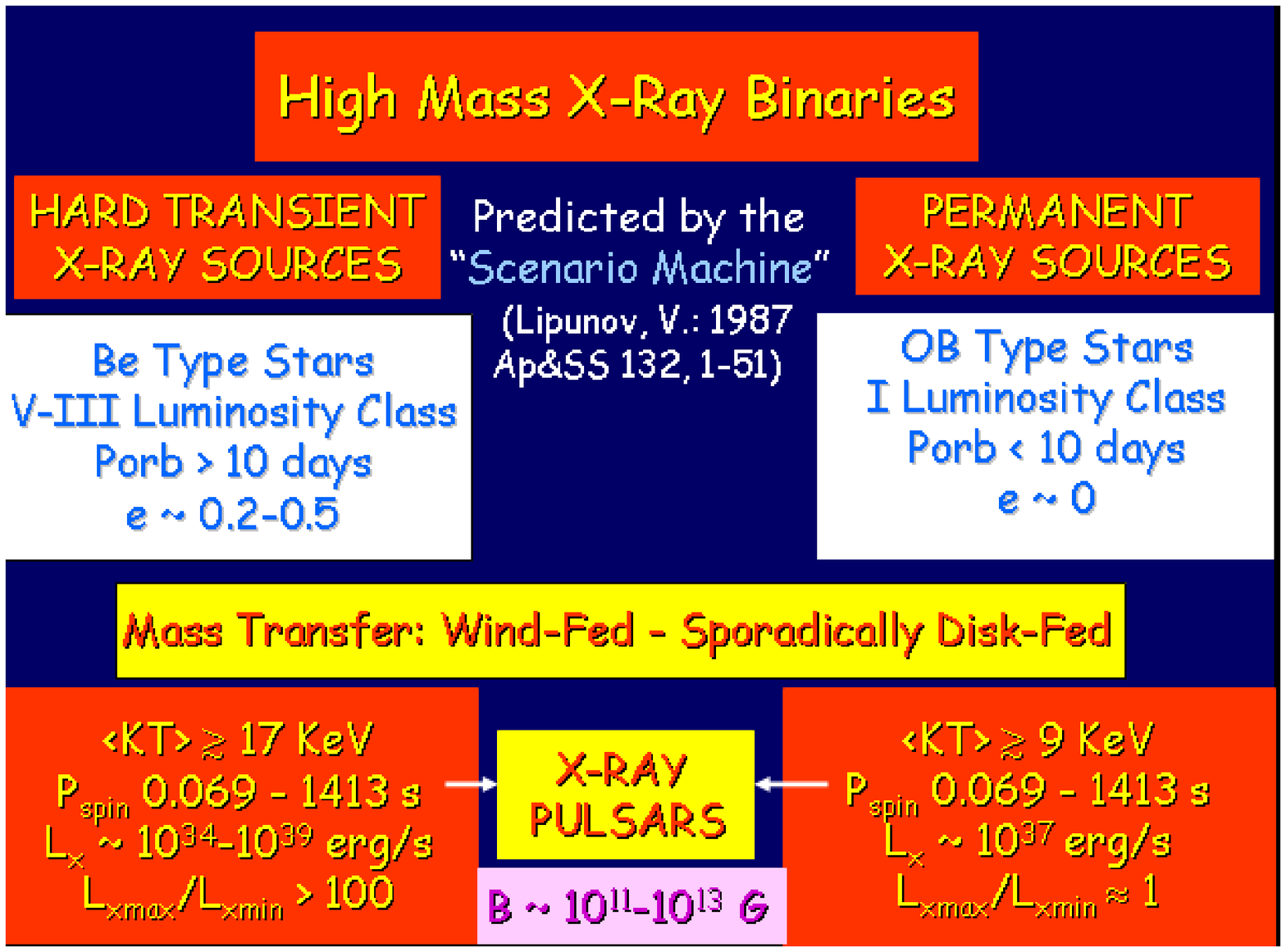}
\caption{Classification of HMXBs (adopted from Giovannelli, 2015).  }
\end{center}
\end{figure*}
%%%%%%%%%%%%%%%%%%%%%%%%%%%%%%%%%%%%%%%%%%%%%%

\subsection{X-ray/Be systems}

The X-ray/Be binaries are the most abundant group of massive X-ray binaries in the galaxy, with a total inferred number of between $10^3$ and $10^4$. The ones which do occasionally flare-up as transient X-ray/Be systems are only the "tip" of this vast "iceberg" of systems (van den Heuvel and Rappaport, 1987). The mass loss processes are due to the rapid rotation of the Be star, the stellar wind and, sporadically, to the expulsion of casual quantity of matter essentially triggered by gravitational effects close to the periastron passage of the neutron star. The long orbital period ($> 10$ days) and a large eccentricity of the orbit ($> 0.2$) together with transient hard X-ray behavior are the main characteristics of these systems. Among the whole sample of galactic systems containing 114 X-ray pulsars (Liu, van Paradijs \& van den Heuvel, 2006), only few of them have been extensively studied. Among these, the system A 0535+26/HDE 245770 -- HDE 245770 was nicknamed Flavia star by Giovannelli \& Sabau-Graziati, 1992) -- is the best known thanks to concomitant favorable causes, which rendered possible forty three years of coordinated multifrequency observations, most of them discussed in the past by e.g. Giovannelli \& Sabau-Graziati (1992), Burger et al. (1996), Piccioni et al. (1999), and later by Giovannelli \& Sabau-Graziati (2011) and Giovannelli et al. (2015a,b).
Accretion powered X-ray pulsars usually capture material from the optical companion via stellar wind, since this primary star generally does not fill its Roche lobe. However, in some specific conditions (e.g. the passage at the periastron of the neutron star) and in particular systems (e.g. A 0535+26/HDE 245770), it is possible the formation of a temporary accretion disk around the neutron star behind the shock front of the stellar wind. This enhances the efficiency of the process of mass transfer from the primary star onto the secondary collapsed star, as discussed by Giovannelli \& Ziolkowski (1990) and by Giovannelli et al. (2007) in the case of A 0535+26.

Optical emission of HMXBs is dominated by that of the optical primary component, which is not, in general,  strongly influenced by the presence of the X-ray source.  The behavior of the primary stars can be understood in the classical (or almost) frame-work of the astrophysics of these objects, i.e. by the study of their spectra which will provide indications on mass, radius, and luminosity.
		Both groups of HMXBs (transient and permanent) differ because of the different origin of the mass loss process: in the first, the mass loss process occurs via a strong stellar wind and/or because of an incipient Roche lobe over-flow; in the  second group, the mass transfer is probably partially due to the  rapid rotation  of  the  primary  star  and partially  to  stellar  wind  and sporadically to expulsions of a casual quantity of matter,  essentially triggered by gravitational effects because of periastron passage  where the effect of the secondary collapsed star is more marked.
		A relationship between orbital period of HMXBs and the spin period of the X-ray pulsars is shown in Fig.  5 (updated from Giovannelli \& Sabau-Graziati, 2001 and from Corbet, 1984, 1986). It allows to recognize three kinds of systems, namely disk-fed, wind-fed [P$_{\rm pulse} \propto$ (P$_{\rm orb}$)$^{4/7}$], and X-ray/Be systems [P$_{\rm pulse} \propto$ (P$_{\rm orb}$)$^2$].

%%%%%%%%%%%%%%%%% FIGURE 5 %%%%%%%%%%%%%%%%%%%%%%%%
\begin{figure*}[!ht] %[!hbp] %[!ht] %[!hbp] %[h!]
\begin{center}
\includegraphics[width=120mm,height=80mm,angle=0] {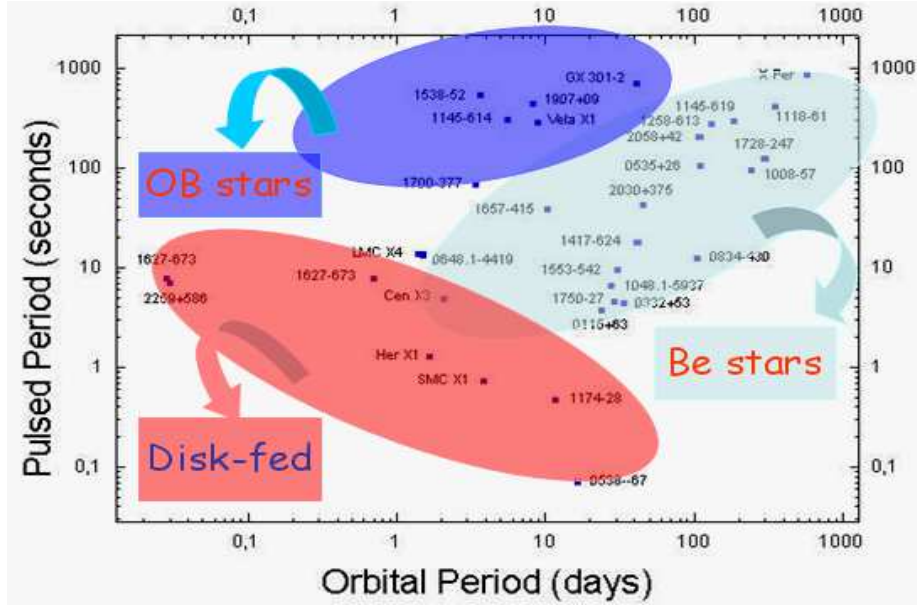}
\caption{Spin period vs orbital period for X-ray pulsars.
Disk--fed systems are clearly separated by systems having as optical
counterparts either OB stars or Be stars (adopted from Giovannelli \& Sabau-Graziati,
2001, after Corbet, 1984, 1986).} %\label{fig4}
\end{center}
\end{figure*}
%%%%%%%%%%%%%%%%%%%%%%%%%%%%%%%%%%%%%%%%%%%%%%%%%%%%

Most of the systems having a Be primary star are hard X-ray (KT $>10$ KeV) transient sources (HXTS). They are concentrated on the galactic plane within a band of $\sim 3.9^\circ$. The orbits are quite elliptic and the orbital  periods large (i.e. A 0538-66: e = 0.7, P$_{\rm orb}$ = 16.6 days (Skinner et al., 1982); A 0535+26: e = 0.47 (Finger, Wilson \& Hagedon, 1994),  P$_{\rm orb}$ = 111.0 days (Priedhorsky \& Terrell, 1983). The X-ray flux during outburst phases is of order 10-1000 times greater than during quiescent phases. For this reason, on the contrary, the stars belonging to the class of permanent X-ray sources, which do not present such strong variations in X-ray emission, can be also named "standard" high mass X-ray binaries.
	In X-ray/Be systems, the primary Be star is relatively not evolved and is contained within its Roche lobe. The strong outbursts occur almost periodically in time scales of the order of weeks-months. Their duration is shorter than the quiescent phases. During X-ray outbursts, spin-up phenomena in several systems have been observed (i.e. A 0535+26 and 4U 1118-61 (Rappaport \& Joss, 1981). The observed spin-up rates during the outbursts are consistent with torsional accretion due to  an accretion disk (e.g. Ghosh, 1994). So, the formation of a temporary accretion disk around the collapsed object should be possible during outburst phases (e.g. Giovannelli \& Ziolkowski, 1990).

%The number of X-ray pulsars slowly increases with time thanks to new detections performed with different new generation observatories. They were 95 in 2000 (Giovannelli \& Sabau-Graziati, 2000) and the orbital periods were known only for about three dozens of them. They contain the group of the permanent HMXBs and that of transient HMXBs (X-ray/Be systems), whose components are an X-ray pulsing neutron star - the secondary - and a giant or supergiant OB or a Be star, respectively - the primary . Moreover, some low-mass X-ray Binaries (LMXBs) containing an X-ray pulsar and some pulsars belonging to Magellanic Clouds were contained too in the sample of 95 systems. In 2006 the known X-ray pulsars were 114 (Liu, van Paradijs \& van den Heuvel, 2006). Coe et al. (2010) report $\sim 60$ X-ray pulsars in the SMC.
%Later, Rajoelimanana \& Charles (2012) listed 49 optical counterparts of SMC X-ray pulsars detected by MACHO and OGLE. Systems with known P$_{\rm spin}$ are 20 while the systems with known P$_{\rm orb}$ are 23, being 6 of them uncertain.

\section{Old \& News from the transient X-ray/Be system A0535+26/HDE245770}

%%%%%%%%%%%%%%%

%%%%%%%%%%%%%%%
The most studied HMXB system, for historical reasons and due to
concomitant favourable causes, is the X-ray/Be system A
0535+26/HDE 245770. By means of long series of coordinated multifrequency
measurements, very often simultaneously obtained, it was possible
to:
\begin{itemize}
  \item identify the optical counterpart HDE 245770 of the
X-ray pulsar;
  \item identify various X-ray outbursts
triggered by different states of the optical companion and influenced
by the orbital parameters of the system;
  \item identify the presence of a temporary
  accretion disc around the neutron star at periastron.
\end{itemize}

Multifrequency observations of A 0535+26 started soon after its
discovery as an X-ray pulsar by the Ariel-5 satellite on April 14, 1975 (Coe et al., 1975).
The X-ray source was in outburst with intensity of $\sim 2$ Crab and showed a pulsation of $\sim 104$ s (Rosenberg et al., 1975). The hard X-ray spectrum during the decay from the April 1975 outburst became softer, so that the 19 May spectrum had $E^{-0.8}$ and the 1 June spectrum $E^{-1.1}$ (Ricketts et al., 1975).
Between 13 and 19 April, 1975, as the nova brightened, the spectra showed some evidence of steepening. The best fit of the experimental data between roughly 27 and 28 April was compatible with an 8 keV black-body curve (Coe et al., 1975). The X-ray source decayed from the outburst with an {\it e}-folding time of 19 days in the energy range of 3-6 keV (Kaluzienski et al., 1975).

In the X-ray error box of the X-ray source A 0535+26, detected by Ariel V, were present 11 stars up to 23$^{rd}$ magnitude and one of them (HDE 245770) of magnitude around 9 showed the H$_\alpha$ and H$_\beta$ in emission, H$_\gamma$ filled in with emisssion, and H$_\delta$, H$_\epsilon$,..., H$_{10}$ in absorption (Margon et al., 1977). A priori probability of finding a 9 mag star in  such a field is 0.004, thus HDE 245770 was considered as the probable optical counterpart of A0535+26.

But in order to
really associate this star with the X-ray pulsar, it was necessary
to find a clear signature proving that the two objects would
belong to the same binary system. This happened thanks to a sudden
insight of one of us (FG), who predicted the fourth X-ray outburst
of A 0535+26 around mid December 1977. For this reason,
Giovannelli's group was observing in optical HDE 245770 around the
predicted period for the X-ray outburst of A 0535+26. Figure 6
shows the X-ray flux intensity of A 0535+26 as deduced by
various measurements available at that time, with obvious
meaning of the symbols used (Giovannelli, 2005). FG's intuition
was sparked by looking at the rise of the X-ray flux (red line) and at
the 24th May 1977 measurement (red asterisk): he assumed that the
evident rise of the X-ray flux would have produced an outburst
similar to the first one, which occurred in 1975. Then with a simple
extrapolation he predicted the fourth outburst, similar to the
second: and this happened!

Optical photoelectric photometry of HDE 245770 showed significant
light enhancement of the star relative to the comparison star BD
+26 876 between Dec. 17 and Dec 21 (here after 771220-E) and successive fading up to Jan. 6
(Bartolini et al., 1978), whilst satellite SAS-3 was detecting an
X-ray flare (Chartres \& Li, 1977). The observed enhancement of
optical emission followed by the flare-up of the X-ray source gave a
direct argument strongly supporting the identification of HDE
245770 -- later nicknamed Flavia' star by Giovannelli \& Sabau-Graziati (1992) -- with A 0535+26.

%%%%%%%%%%%%%%%%% FIGURE 6  %%%%%%%%%%%%%%%%%%%%%%%%
\begin{figure*}[!ht]
\begin{center}
\includegraphics[width=120mm,height=85mm,angle=0] {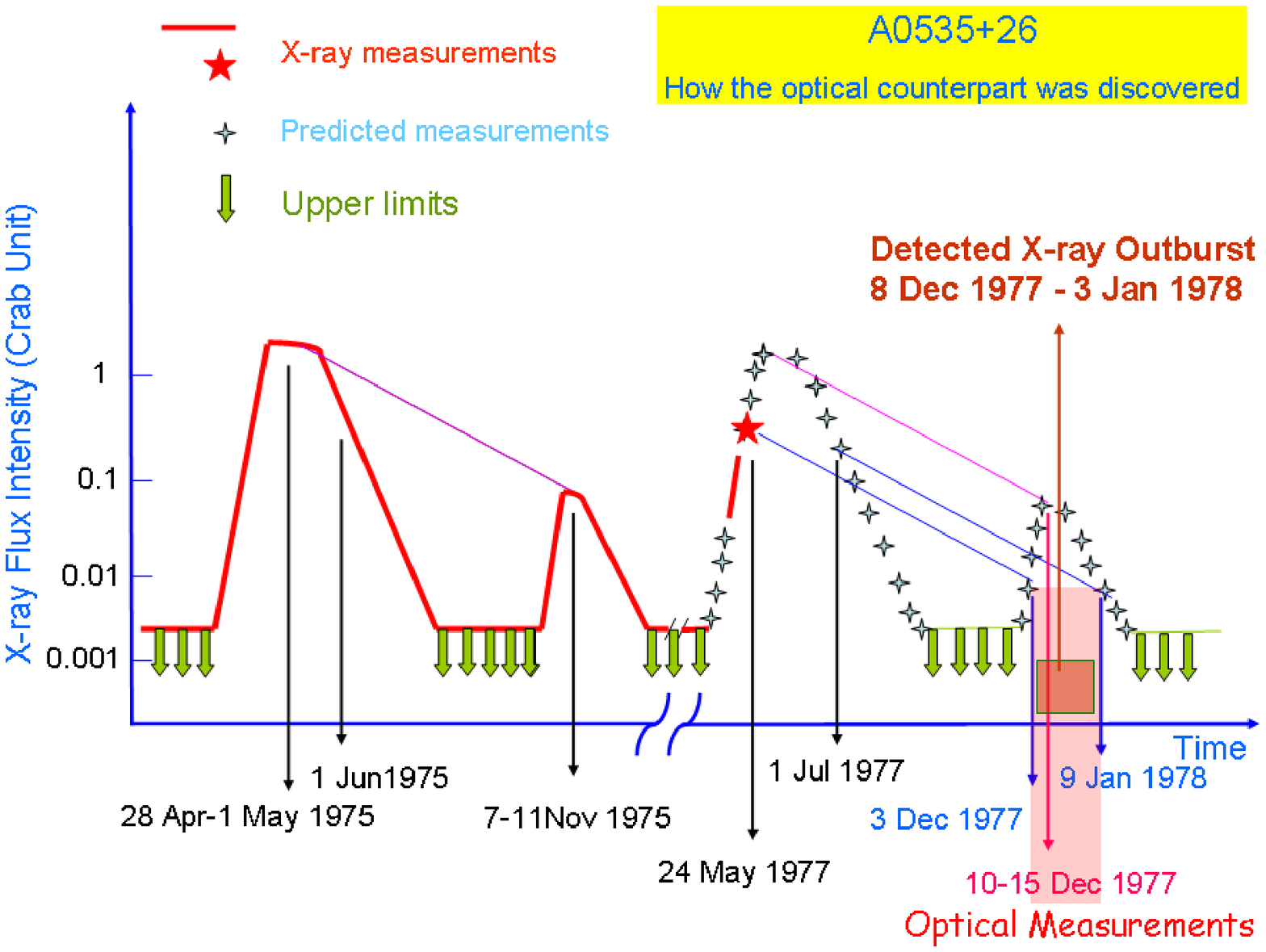}
\caption{X-ray flux versus time of A 0535+26. X-ray measurements
are reported with red lines and asterisk, upper limits with green
arrows, and predicted fluxes with light blue stars. Periods of
real detected X-ray outburst and optical measurements are also marked (adopted from Giovannelli, 2005).}
\end{center}
\end{figure*}
%%%%%%%%%%%%%%%%%%%%%%%%%%%%%%%%%%%%%%%%%%%%%%%%%%%%

Soon after, with spectra taken at the Loiano 152 cm telescope with a
Boller \& Chivens 26767 grating spectrograph (831 grooves/mm
II-order grating: 39 \AA\, mm$^{-1}$) onto Kodak 103 aO plates, it
was possible to classify HDE 245770 as O9.7IIIe star. This
classification was so good that it survives even to the recent
dispute attempts made with modern technology. The mass and
radius of the star are 15 M$_\odot$ and 14 R$_\odot$,
respectively; the distance to the system is $1.8 \pm 0.6$ kpc
(Giangrande et al., 1980).

UV spectra taken with the IUE enabled the reddening
of the system to be determined as E(B-V) = $0.75 \pm 0.05$ mag, the rotational
velocity of the O9.7IIIe star (v$_{\rm rot}\sin i = 230 \pm 45$ km
s$^{-1}$), the terminal velocity of the stellar wind (v$_\infty
\simeq 630$ km s$^{-1}$), the mass loss rate (\.{M} $\sim 10^{-8}$
M$_\odot$ yr$^{-1}$ in quiescence (Giovannelli et al., 1982).
During the October 1980 strong outburst, the mass loss rate was
\.{M} $\sim 7.7 \times 10^{-7}$ M$_\odot$ yr$^{-1}$ (de Martino et
al., 1989).

Complete reviews of this system can be found in Giovannelli et al. (1985), Giovannelli \& Sabau-Graziati
(1992), and Burger et al. (1996).

Briefly, the properties of this system, placed at a distance of $1.8 \pm 0.6$ kpc (Giangrande et al., 1980), can be summarized as follows: a hard X-ray transient, long-period X-ray pulsar -- the secondary star -- is orbiting around the primary O9.7IIIe star. The masses are of $\sim 1.5 \pm 0.3$ M$_\odot$ (Joss \& Rappaport 1984; Thorsett et al. 1993; van Kerkwijk, van Paradijs, J. \& Zuiderwijk, 1995), and 15 M$_\odot$ (Giangrande et al., 1980) for the secondary and primary star, respectively. The eccentricity is e = 0.47 (Finger et al., 1994). Usually the primary star does not fill its Roche lobe (de Loore et al., 1984).
However, the suggestion that there might be a temporary accretion disk around the X-ray pulsar when it approaches periastron (Giovannelli \& Zi\'{o}{\l}kowski, 1990) was confirmed by the X-ray measurements of Finger, Wilson \& Harmon (1996) and was discussed by Giovannelli et al. (2007).

The first suggestion of Bartolini et al. (1983) about the value of the orbital period (P$_{\rm orb} = 110.856 \pm 0.002$ days), allowed Giovannelli \& Sabau-Graziati (2011) to discover a systematic delay ($\sim$ 8 days) of the X-ray outbursts with respect to the periastrons passages of the neutron star. Just a little before or simultaneously to the periastron, the system experiences an optical brightening ranging from $\approx 0.02$ to $\approx 0.2$ magnitudes.

The trigger of this discovery was the optical flare occurred at JD 2,444,944 (5th December 1981) (hereafter 811205-E; E stands for event) (Giovannelli et al., 1985) and followed by a short X-ray outburst (811213-E) (Nagase et al., 1982) -- predicted by a private communication of Adriano Guarnieri (member of Giovannelli's group) to the team of Hakucho Japanise X-ray satellite.
Unfortunately in the following years simultaneous optical and X-ray measurements were not always obtained around  periastron passage. However, the available data  were sufficient for showing the aforementioned systematic delay.

In order to  explain and describe the 8-day delay between periastron passage and X-ray outbursts, GBK13 constructed  a model  adopting  the orbital period determined by Priedhorsky \& Terrell (1983) from X-ray data (P$_{\rm orb} = 111.0 \pm 0.4$ days), and the ephemeris JD$_{\rm opt-outb}$ = JD$_0$(2,444,944) $\pm$ n(111.0 $\pm$ 0.4) days; the 111-day orbital period agrees within the error bars with many other determinations reported in the literature (from optical data, e.g. Guarnieri et al., 1985; de Martino et al., 1985; Hutchings, 1984; Janot-Pacheco, Motch \& Mouchet, 1987). From X-ray data (e.g. Nagase et al., 1982; Priedhorsky \& Terrell, 1983; Motch et al., 1991; Finger, Wilson \& Harmon, 1996; Coe et al., 2006; Finger et al., 2006).

In the following we will discuss the experimental results that corroborate GBK13 model, by the analysis of several results coming from multifrequency LE and HE observations of A 0535+26/HDE 245770 nicknamed Flavia' star.

\section{Time delay between optical and X-ray outbursts in A 0535+26/HDE 245770}

A description of the time--delay among many optical and X-ray events occurring around the periastron passages in the system A 0535+26/HDE 245770 have been presented in the papers by GBK13 and Giovannelli et al. (2015a,b). However, just to remark the importance of the experimental evidence of such a time--delay we will present a few more examples that in our opinion definitively support the validity of the model developed in GBK13. Briefly, the model is the following:
in the vicinity of periastron the mass flux $\dot{\rm M}$ increases (depending on the activity of the Be star) between  $\approx 10^{-8}$ and $\approx 10^{-7}$ M$_\odot$ yr$^{-1}$. The outer part of the accretion disk -- geometrically thin and optically thick without advection (Shakura \& Sunyaev, 1973; Bisnovatyi-Kogan, 2011) -- becomes hotter, therefore the optical luminosity (L$_{\rm opt}$) increases. Due to large turbulent viscosity, the wave of the large mass flux is propagating toward the neutron star, thus the X-ray luminosity (L$_{\rm x}$) increases due to the appearance of a hot accretion disk region and due the accretion flow channeled by the magnetic field lines onto magnetic poles of the neutron star. The time--delay $\tau$ is the time between the optical and X-ray flashes appearance. Figure 7 shows a sketch of this model.

%%%%%%%%%%%%%%%%%%%% Figure 7 %%%%%%%%%%%%%%
\begin{figure}[!ht]%[!hbp]%[!ht]%[!hbp]%[h][h!]
\centering
\includegraphics[width=0.8\textwidth]{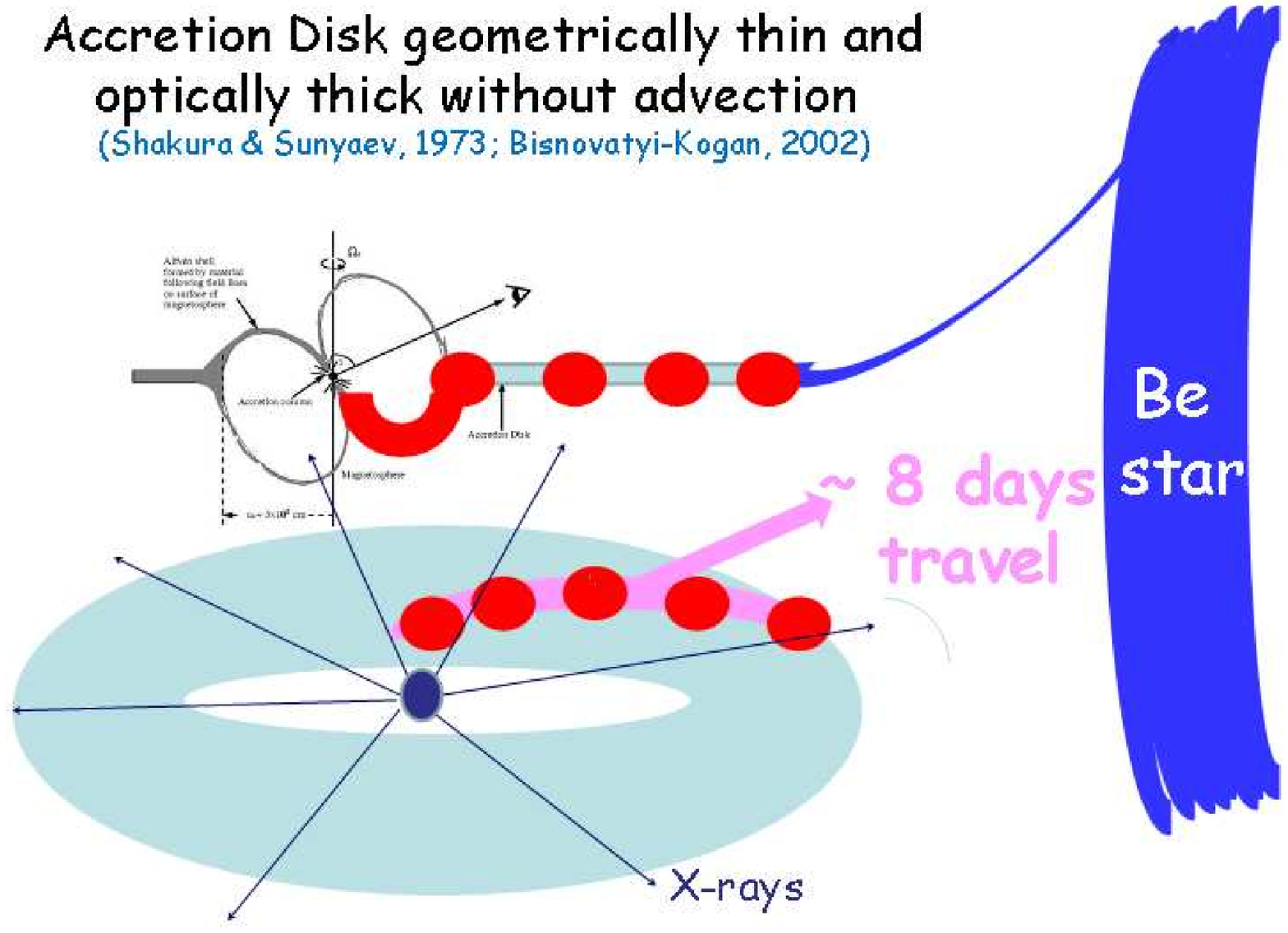}
\caption{Sketch of the viscous accretion disk model for explaining the time-delay between X-ray and optical flashes (adopted from GBK13). }
\label{fig1b}
\end{figure}
%%%%%%%%%%%%%%%%%%%%%%%%%%%%%%%%%%%%%%%%%%%%

\noindent
By using the ephemerides given by GBK13, namely:

JD$_{\rm opt-outb}$ = JD$_0$(2,444,944) $\pm$ n(111.0 $\pm$ 0.4) days

\noindent
that fixed the reference point at the date 5th December 1981 (811205-E),
it was possible to explain the behaviour of the system during the year 2014 (Giovannelli et al., 2015a). It was possible not only to predict the arrival time of the X-ray outbursts following the optical flashes, but also the intensity I$_{\rm x}$ of the X-ray flares, thanks to the relationship I$_{\rm x}$ versus $\Delta$V$_{\rm mag}$, where $\Delta$V$_{\rm mag}$ is the relative variations of the V magnitude of the Be star around the periastron passage with respect to the level before and after such a passage. This relationship is shown in Fig. 8 (adapted from Giovannelli et al., 2015b).

%%%%%%%%%%%%%%%%%%%% Figure 8 %%%%%%%%%%%%%%
\begin{figure}[!ht]%[!hbp]%[!ht]%[!hbp]%[h][h!]
\centering
\includegraphics[width=0.6\textwidth]{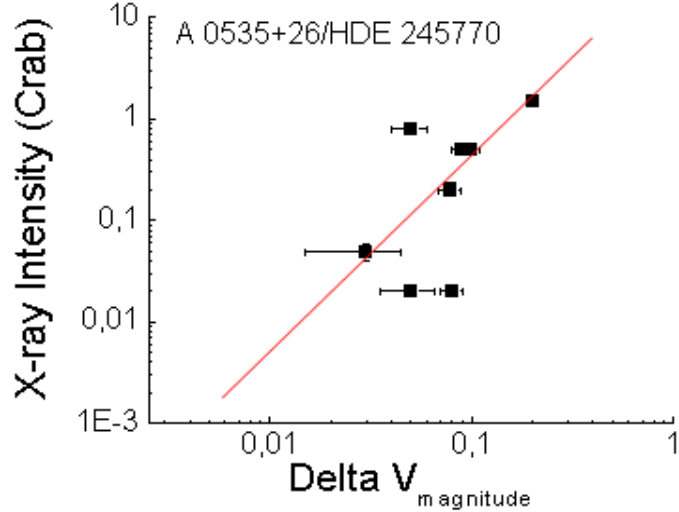}
\caption{Intensity of the X-ray flare of A 0535+26 versus the variation of V magnitude of HDE 245770 around the periastron passage (adapted from Giovannelli et al., 2015b). }
\label{fig2}
\end{figure}
%%%%%%%%%%%%%%%%%%%%%%%%%%%%%%%%%%%%%%%%%%%%

We have also found a relationship between the equivalent width (EW) of H$_\alpha$ and I$_{\rm x}$. The values of H$_\alpha$-EW have been taken only if measurements were performed around the periastron passage $\pm 10$ days. Unfortunately few measurements have been found in this time range. However, the trend of the relationship is rather good, as shown in Fig. 9, where data taken from Camero-Arranz et al. (2012), Yan, Li \& Liu (2012),
Giovannelli et al. (2015b) are reported (Fasano, 2015).
It is interesting to note that in one occasion, at the 106th periastron passage (JD 2,456,710 = 21 Feb 2014) after 811205-E, optical photometry and spectroscopy as well as X-ray measurements from different experiments were obtained. A jump in the H$_\alpha$-EW and H$_\beta$-EW in correspondence with the rise of X-ray intensity was detected, being the jump of H$_\beta$-EW delayed of $\approx$ 5 days with respect to that of H$_\alpha$-EW. This important result deserves further investigations. However, the jumps of H$_\alpha$-EW and H$_\beta$-EW could originate because of a contribution to the total emission in those lines coming from the temporary accretion disk around the neutron star (Giovannelli et al., 2015a, and the references therein). And if so, the delay between H$_\beta$-EW and H$_\alpha$-EW jumps should be explicable within the framework of GBK13's model.

%%%%%%%%%%%%%%%%%%%% Figure 9 %%%%%%%%%%%%%%
\begin{figure}[!ht]%[!hbp]%[!ht]%[!hbp]%[h][h!]
\centering
\includegraphics[width=0.6\textwidth]{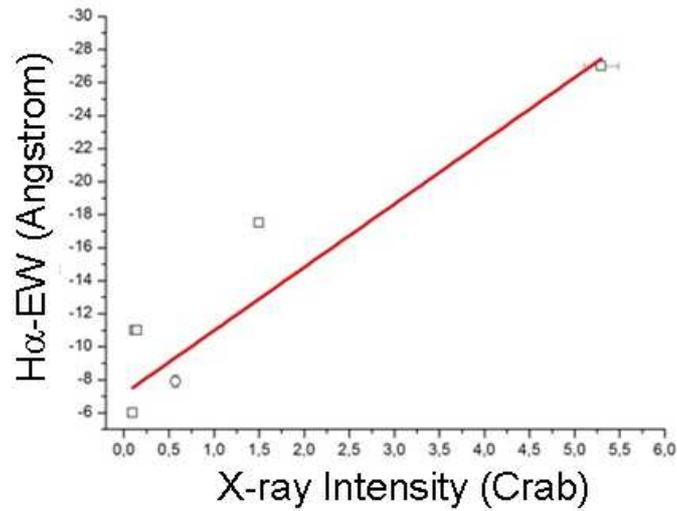}
\caption{Intensity of the X-ray flare of A 0535+26 versus the equivalent width of H$_\alpha$ of HDE 245770 around the periastron passage (data taken from Camero-Arranz et al., 2012; Yan, Li \& Liu, 2012; Giovannelli et al., 2015b) (figure adopted from Fasano, 2015). }
\label{fig3}
\end{figure}
%%%%%%%%%%%%%%%%%%%%%%%%%%%%%%%%%%%%%%%%%%%%

An impressive strong optical event have been detected on March 19, 2010 (JD 2,455,275).
On the basis of such a strong optical activity -- especially H$_\gamma$ in emission -- Giovannelli, Gualandi \& Sabau-Graziati (2010, ATel 2497) predicted the incoming X-ray outburst of A 0535+26, which actually occurred (Caballero et al., 2010b, ATel 2541). The X-ray intensity reached was 1.18 Crab on April 3, 2010 in the range 15--50 keV of BAT/SWIFT (Caballero et al., 2010a,b,c,d; Caballero et al., 2011). Figure 10 shows the March--April 2010 event. The X-ray flare started about 8 days after the 93th periastron passage after the 811205-E, just when optical spectroscopy was performed by Giovannelli, Gualandi \& Sabau-Graziati (2010), and reached the maximum about 12 days later and decayed in about 20 days roughly as occurred in 1975 when A0535+26 was discovered by the Ariel V satellite.

%%%%%%%%%%%%%%%%%%%% Figure 10 %%%%%%%%%%%%%%
\begin{figure}[!ht]%[!hbp]%[!ht]%[!hbp]%[h][h!]
\centering
\includegraphics[width=0.7\textwidth]{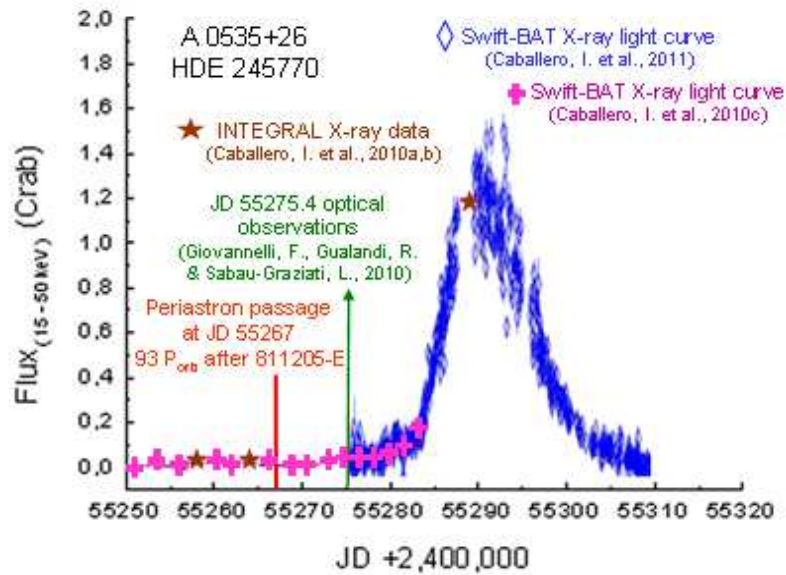}
\caption{The predicted March--April 2010 X-ray outburst of A 0535+26 (Giovannelli, Gualandi \& Sabau-Graziati, 2010) after the 93th passage at the periastron after 811205-E (Caballero et al., 2010a,b,c,d; Caballero et al., 2011).  }
\label{fig4}
\end{figure}
%%%%%%%%%%%%%%%%%%%%%%%%%%%%%%%%%%%%%%%%%%%%

The astonishing fact that definitively demonstrate the goodness of GBK13's ephemerides and the mechanism triggering the X-ray outburst with a time--delay with respect to the optical flare around the periastron passage is reported in Fig. 11.

%%%%%%%%%%%%%%%%%%%% Figure 11 %%%%%%%%%%%%%%
\begin{figure}%[!ht]%[!hbp]%[!ht]%[!hbp]%[h][h!]
\centering
\includegraphics[width=0.7\textwidth]{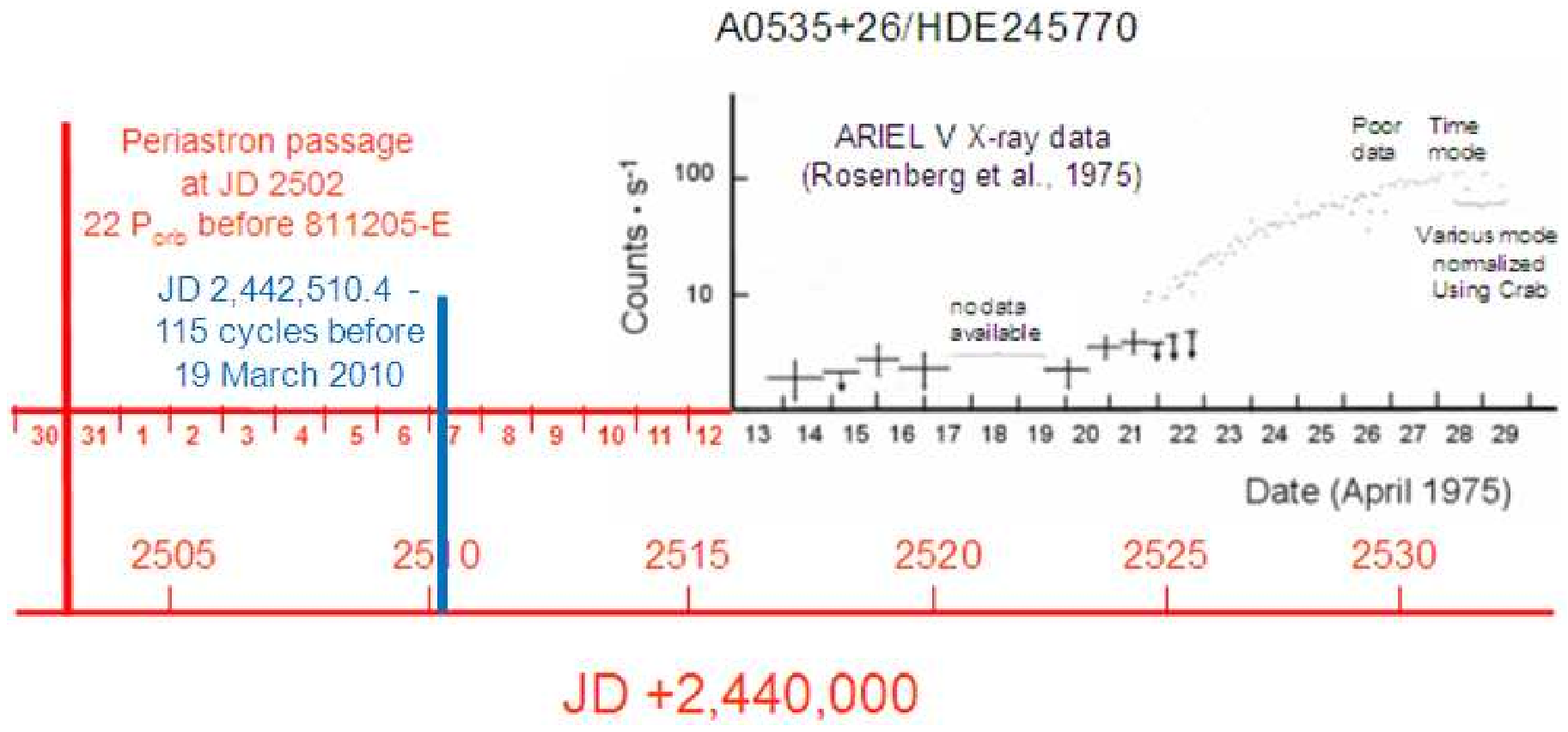}
\caption{The Periastron passage at the 22nd cycle before 811205-E (JD 2502) (red line) precedes of $\sim 14$ days the X-ray outburst of A 0535+26 which starts approximatively on JD 2516 (after Rosenberg et al., 1975). The vertical blue line indicates the day April 7, 1975 (JD 2,442,510.4 ) just 115 cycles before the strong optical spectroscopic activity detected on March 19, 2010 (JD 2,455,275.4) (after GBK13).  }
\label{fig6}
\end{figure}
%%%%%%%%%%%%%%%%%%%%%%%%%%%%%%%%%%%%%%%%%%%%

Indeed, in Fig. 11 the measurements of the first detection of A 0535+26 by Ariel V satellite are reported. Unfortunately in 1975, around the time of the discovery of A 0535+26 no optical measurements are available for obvious reasons. The vertical red line indicates the time of the periastron passage following GBK13's ephemeris. This passage occurred at the 22nd cycle before the 811205-E.
The vertical blue line indicates the day April 7, 1975 (JD 2,442,510.4 ) just 115 cycles before the strong optical spectroscopic activity detected on March 19, 2010 (JD 2,455,275.4), that preceded the strong X-ray outburst reported in Fig. 10.

The similarity between the first X-ray outburst and that of March--April 2010 is evident, and the separation of the two events is exactly 115 cycles.

\section{General model of time lag between optical and X-ray outbursts in binary accreting sources}

%\subsection{Accretion onto Neutron Stars and Black Holes}

In LMXBs (Low-Mass X-ray Binaries) the compact object can be either a neutron star or a black hole and the optical companion is a low mass star. The exchange of matter occurs via Roche lobe overflow, like shown in the left panel of Fig. 3. In HMXBs (High-Mass X-ray Binaries) the compact object can be either a neutron star or a black hole and the optical companion is a high mass star: giant or super-giant. The exchange of matter occurs mainly via stellar wind since usually the optical star does not fill its Roche lobe (Fig. 3, left panel). However, sometimes, the exchange of matter can occur in a mixed way because of the formation of an accretion disk around the compact object around the periatron passage (e.g. Giovannelli \& Zi\'{o}{\l}kowski, 1990) (Fig. 3, right panel).

McClintock, Narayan \& Rybicki (2004) found a very interesting relationship between the minimum X-ray luminosity in the range 0.5--10 keV and the orbital periods of BH LMXBs and NS LMXBs.
BH LMXBs are on average a factor of $\sim$ 100 fainter than NS LMXBs with similar orbital periods (Fig. 12).

%%%%%%%%%%%%%%%%%% FIGURE 12 %%%%%%%%%%%%%%%%%
\begin{figure*}%[!hbp]%[!ht]%[!hbp]%[!ht]%[!hbp]%[h]
\begin{center}
\includegraphics[width=10cm]{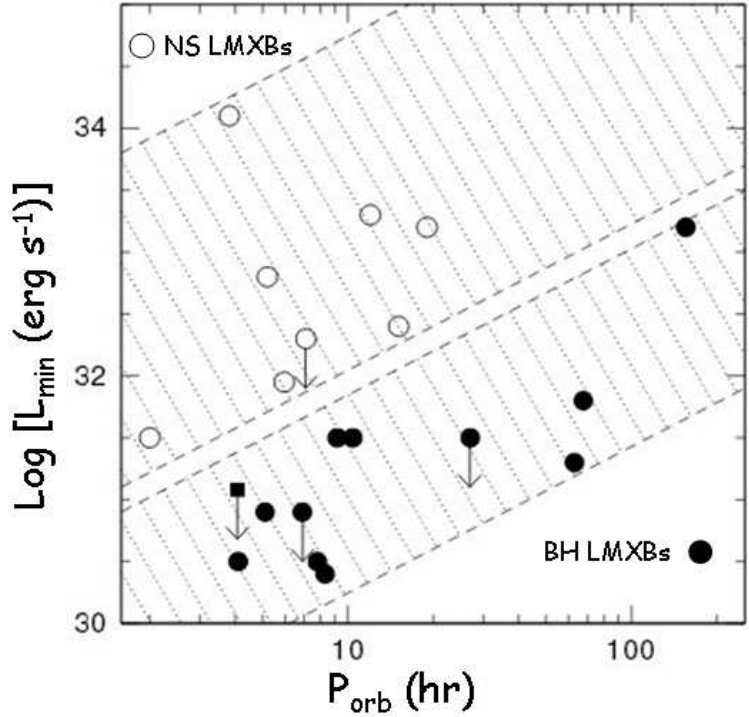}
\caption{Minimum X-ray luminosity (0.5-10 keV) versus orbital period for BH LMXBs and NS LMXBs. The diagonal hatched areas delineate the regions occupied by the two classes of sources and indicate the dependence of luminosity on orbital period (adapted from McClintock, Narayan \& Rybicki, 2004). }
\end{center}
\end{figure*}
%%%%%%%%%%%%%%%%%%%%%%%%%%%%%%%%%%%%%%%%%%%%%%

As well known, X-ray/Be systems are formed by a compact star and an optical star. Obviously there is a mutual influence between the two stars. Low-energy (LE) processes influence high-energy (HE) processes and vice versa. Never confuse the effect with the cause. There is a general law in  the Universe: {\bf Cause and Effect}. The {\it Cause} generates an {\it Effect} and NOT vice versa!

%Time-lag between HE events and LE events in disk-fed accreting X-ray binaries (XRBs) has been noted in many systems, but the trigger of the work resulted in a model for explaining in general such a phenomenon (Bisnovatyi-Kogan \& Giovannelli, 2017) was given by Giovannelli \& Sabau-Graziati (2011) who noted a systematic delay between the relative enhancement in luminosity of the optical Be star -- occurring at the periastron passage of the neutron star -- and the subsequent X-ray flare in the system HDE 245770/A 0535+26. The model for such a system was developed  and corroborated by many events (Giovannelli, Bisnovatyi-Kogan \& Klepnev, 2013: GBK13), and later by events reported in Giovannelli et al. (2015) where also a relationship between $\Delta$V$_{\rm mag}$ of the optical star at the periastron and X-ray intensity (I$_{\rm X}$) of the 8-day delayed flare was produced.

%The time--delay $\tau$ is the time between the optical and X-ray flashes appearance.

%%%%%%%%%%%%%%%%% da South Africa 2016 %%%%%%%%%%%
It is right to remind that the mechanism proposed by GBK13 for explaining the X-ray-optical delay in A 0535+26/HDE 245770 is based on an enhanced mass flux propagation through the viscous accretion disk. This mechanism, known as UV-optical delay (the delay of the EUV flash with respect to the optical flash) was observed and modeled for cataclysmic variables (e.g. Smak, 1984; Lasota, 2001). Time delays have been detected also in several other X-ray transient binaries. This is the reason that urged Bisnovatyi-Kogan \& Giovannelli (2017, BKG17) to generalize the aforementioned model, developed for the particular case of A 0535+26/HDE 245770 (Flavia' star). This general model provides the formula (6.1) of the time delay between the optical and X-ray flashes appearance in transient cosmic accreting sources:

\begin{equation}
\tau=6.9 \frac{{\rm m}^{^2/_3} \dot{{\rm m}}^{^1/_{15}}}{\alpha^{^4/_5}\left( {\rm T}_{4} \right)^{^{28}/_{15}}}
\label{eq:formula}
\end{equation}

%\begin{align*}
\noindent
where:

\smallskip
\noindent
m = M/M$_\odot$\hspace{0.1cm}; \hspace{1.0cm} $\dot{\rm m} = \dot{\rm M}$/(10$^{-8}$ M$_\odot$/yr)\hspace{0.1cm};
\hspace{1.0cm} T$_4 = {\rm T}_0/10^4$ K\hspace{0.1cm}; \hspace{1.0cm}  $\alpha$ = viscosity\hspace{0.1cm}, and\\
T$_0$ = maximum temperature in optics.

%\end{align*}
\bigskip
\noindent
By using this formula it is possible to obtain an excellent agreement between the experimental and theoretical delays found in:

\begin{itemize}
  \item X-ray/Be system A0535+26/HDE245770: $\tau_{\rm exp} \simeq$ 8 days (GBK13); $\tau_{\rm th} \simeq$ 8 days;
  \item Cataclysmic variable SS Cygni; $\tau_{\rm exp}$ = 0.9--1.4 days (Wheatley, Mauche \& Mattei, 2003); $\tau_{\rm th} \simeq$ 1.35 days;
  \item Low-mass X-ray binary Aql X-1/V1333 Aql: $\tau_{\rm exp} \sim$ 3 days (Shahbaz et al., 1998); $\tau_{\rm th} \simeq$ 3.2days
  \item Black hole X-ray transient GRO J1655-40: $\tau_{\rm exp} \sim 6$ days (Orosz et al., 1997); $\tau_{\rm th} \simeq$ 6.5 days.
\end{itemize}

In this general formula the $\alpha$-viscosity parameter plays an important role, and usually it is hard to be determined. However, if the other parameters are known, because experimentally determined, the formula (6.1) can be used for determining $\alpha$, taking into account the experimental delay measured in a certain source.

Over the last couple of decades we have witnessed the discovery of a multitude of
highly ionized absorbers in high-resolution X-ray spectra from both BH and NS XRBs.
The first detections were obtained thanks to ASCA on the BH binaries GROJ1655-40 and GRS 1915+105. Narrow absorption lines in the spectra of these systems identified as Fe XXV and Fe XXVI indicated the first of many discoveries of photo-ionized plasmas in LMXBs (Chandra, XMM-Newton and Suzaku).
Black hole hot accretion flows occur in the regime of relatively low accretion rates and are operating in the nuclei of most of the galaxies in the universe.
One of the most important progress in recent years in this field is about the wind or outflow.
This progress is mainly attributed to the rapid development of numerical simulations of accretion flows, combined with observations on, e.g., Sgr A$^\star$, the SMBH in the Galactic center.
The mass loss from a BH via wind is related to the mass accretion rate onto the BH as (Yuan, 2016):
\bigskip
\begin{center}
$\dot{\rm M}_{\rm wind}$(r) = $\dot{\rm M}_{\rm BH}$ $\times$ (r/20r$_{\rm g})^{\rm s}$ \hspace{1cm} with \hspace{0.3cm} s $\approx$ 1 \hspace{0.5cm} and \hspace{0.5cm} ${\rm r_g} = \frac{2{\rm GM_{BH}}}{{\rm c}^2}$ \hspace{1cm} (6.2)
\end{center}
\bigskip
\noindent

At this point it is useful to make a sort of summary about the number of XRBSs, including CVs.
Liu, van Paradijs \& van den Heuvel (2006, 2007) and Zi\'{o}{\l}kowski (2013) report 315 galactic XRBs: 197 LMXBs (63\%) and 118 HMXBs (37\%), 72 of which are Be/X-ray systems; moreover there are 62 BH candidates. Coleiro \& Chaty (2013) report that in the Milky way there are $\geq$ 200 HMXBs. Ritter \& Kolb (2003) catalogue, in the 7.20 (Dec. 2013) version, reports 1166 CVs.
Buckley (2015) reports about the discoveries of 530 new CVs from MASTER-Network and 855 CVs from Catalina Real Time Survey (CRTS) (http://nesssi.cacr.caltech.edu/DataRelease/). Ferrario, de Martino \& G\"{a}nsicke (2015) report the number of MCVs as $\approx$ 250, and $\sim$ 60 of which IPs, and considering those systems for which the magnetic field intensity has not yet been determined, their number is of $\approx$ 600.
Table 1 shows the content of XRBSs, including CVs, in the Galaxy, and in the LMC and SMC (Zi\'{o}{\l}kowski, 2013; Ferrario, de Martino \& G\"{a}nsicke, 2015; Buckley, 2015). The mass are expressed in unit of SMC.

%%%%%%%%%%%%%%%%%%%%% Table 1 %%%%%%%%%%%%%%%%%%%%%%%
%\footnotesize
\begin{table}[!hbp]
\begin{center}
\caption{Comparison of numbers of different classes
of X-ray Binary Systems in the Milky Way and in the Magellanic Clouds (Zi\'{o}{\l}kowski, 2013; Ferrario, de Martino \& G\"{a}nsicke, 2015; Buckley, 2015).}
\bigskip
\begin{tabular}{|l|c|c|c|l|}
\hline
   &   &  & \\
Name of the Class & Milky Way   & LMC & SMC \\
        &        &       &    \\
%   &   &  \\
\hline
%   &   &  \\
Total mass of the galaxy &    &  & \\
(in M$_{\rm SMC}$ units)  & 100 & 10 & 1\\
\hline
High Mass X-ray Binaries & 118  & 26 & 83\\
in this Be/X-ray & 72 & 19 & 79\\
\hline
Low Mass X-ray Binaries & 197  & 2 & -   \\
\hline
Black Hole Candidates   &  62   & 2 & -  \\
\hline
Cataclysmic Variables & $\approx$ 2000 & - & - \\
in this MCVs & $\approx$ 250 & - & - \\
IPs & $\sim$ 60 & - & - \\
B not yet determined & $\approx$ 600 & - & - \\
\hline
\end{tabular}
\end{center}
\end{table}
%\end{footnotesize}

Grimm (2003) published: (i) a list of the 17 most luminous LMXBs contributing $\approx$ 90\% to the integrated luminosity of LMXBs in the 2-10 keV band in the whole Galaxy, averaged over 1996-2000. The 12 most luminous sources (Cir X-1, GRS 1915+105, Sco X-1, Cyg X-2, GX 349+2, GX 17+2, GX 5-1, GX 340+0, GX 9+1, NGC 6624, Ser X-1, GX 13+1) contribute $\approx$ 80\% of the integrated luminosity of the Galaxy; (ii) a list of the 10 most luminous HMXBs
(Cyg X-3, Cen X-3, Cyg X-1, X 1657-415, V 4641 Sgr, GX 301-2, XTE J1855-024, X 1538-522, GS 1843+009, X 1908+075-
that contribute $\approx$ 40\% to the integrated luminosity of HMXBs in the 2-10 keV band in the whole Galaxy, averaged over 1996-2000.

Raguzova \& Lipunov (1999) -- using the "Scenario Machine" developed by Lipunov (1987) and Lipunov \& Postnov (1988) -- obtained an evolutionary track that can lead to the formation of Be/BH systems. This result has been confirmed fifteen years later by Casares et al. (2014) who discovered MWC 656, the first Be/BH binary.

Indeed, Raguzova \& Lipunov (1999) calculations show that binary black holes with Be stars must have 0.2 < e < 0.8.
It is particularly difficult to detect such systems as most of their spectroscopic variations occur in a relatively small portion of the orbit, and could easily be missed if the systems are observed at widely separated epochs.
This represents one more reason for asking continuous multifrequency observations of different classes of cosmic sources in order to understand their true behaviour.

The critical initial mass of the supernova star that collapses to a BH is accepted to be equal to 55 < M$_{\rm cr} < 75$ M$_\odot$, and the fraction of the presupernova mass (M$_\star$) collapsing to the BH, k$_{\rm BH} =$ M$_{\rm BH}$/M$_\star$ = 0.5. The kick velocity v$_{\rm m}$ = 0--200 km s$^{-1}$.
The age of the system, according to their evolutionary scenario is $4\times10^6$ yr.

The expected number of Be/BH binaries -- with orbital period 10 d < P$_{\rm orb}$ < 1000 d, and eccentricity  0.2 < e < 0.8 -- is 1 Be/BH  for  20-30 Be/NS.

Belczynski and Zi\'{o}{\l}kowski (2009) used binary population synthesis models to show that the expected ratio of Be/XRBs with neutron stars to black holes in the Galaxy is relatively high ($\sim 30-50$), and so broadly in line with observations. Thus we can expect 1 Be/BH  for  30--50 Be/NS.

Therefore, we can expect 1 Be/X-ray BH system for 20--50 Be/X-ray NS systems (Raguzova \& Lipunov, 1999; Belczynski \& Zi\'{o}{\l}kowski, 2009). We know 60 Be/X-ray NS systems (after INTEGRAL).
Thus we expect 1--3 Be/X-ray BH systems. One of this systems has been detected: MWC 656 (Casares et al., 2014).

New simulations -- using the StarTrack binary population synthesis models have been conducted to understand the formation channel of MWC 656 -- constrain the population of Be/BH systems and study the fate of MWC 656 as a possible NS--BH merger, and then possible gravitational wave emitter. In particular, it has been assumed that all donors beyond main sequence are allowed to survive the common envelope (CE) phase. Ten Gyr of evolution of the Galactic disk originates $\sim 8700$ B/BH systems, and 1/3 of them would be Be/BH systems: namely $\sim 2900$).
However, only 13 of them had periods, eccentricities and masses similar to MWC 656 (Grudzinska et al., 2015).

%%%%%%%%%%%%%%%%%% http://hubblesite.org/explore_astronomy/black_holes/encyclopedia.html %%%%%%%%%%%%%
There are so many black holes in the Universe that it is impossible to count them.

Stellar-mass black holes (SBHs) form from the most massive stars when their lives end in supernova explosions. The Milky Way galaxy contains some 10$^{11}$ stars. Roughly one out of every thousand stars that form is massive enough to become a black hole. Therefore, our galaxy must harbor some 10$^8$ stellar-mass black holes. Most of these are invisible to us, and only nineteen have been identified (Wiktorowicz, Belczynski \& Maccarone, 2014) with masses up to $\sim 15$ M$_\odot$. Theoretically the mass of a SBH depends on the initial mass of the progenitor, how much mass is lost during the progenitor's evolution and on the supernova explosion mechanism (Belczynski et al., 2010; Fryer et al., 2012). Mass is lost through stellar winds, and the amount of mass lost strongly depends on the metallicity of the star. For a low metallicity star ($\sim 0.01$ of the solar metallicity) it is possible to leave a black hole of $\leq 100$ M$_\odot$ (Belczynski et al., 2010).
In the region of the Universe visible from Earth, there are perhaps 10$^{11}$ galaxies. Each one has about 10$^8$ stellar-mass black holes. And somewhere out there, a new stellar-mass black hole is born in a supernova every second.

However, some attempts of evaluation of the number of SBHs in the Galaxy have been done. For instance, taking into account the $\gamma$-ray emissivity of the Galaxy (1.3 $\times 10^{43}$ s$^{-1}$ for E $>$ 100 MeV) measured by the SAS II satellite  (Strong, Wolfendale \& Worral, 1976) and the processes of disk-fed accretion onto black holes, Giovannelli, Karaku{\l}a \& Tkaczyk (1981, 1982) -- considering a spherical accretion flow  with  a constant Mach number, corresponding to the adiabatic power $\gamma$ = 5/3 -- found a possible upper limits to the number of black holes (M $\sim$ 10 M$_\odot$ and $\dot{\rm M} \approx 10^{-8}$ M$_\odot$ yr$^{-1}$) of 10$^{-5}$ - 10$^{-4}$ of the total star population of the Galaxy.
%for Mach's number 1 and 2, respectively.
%$^{(\ast)}$

There is a class of intermediate-mass black holes (IMBHs), with masses $> 100$ M$_\odot$ up to $\approx 10^5$ M$_\odot$. It contains a dozen systems, as listed in Johnstone (2004). However, black holes with masses of several hundred to a few thousand solar masses remain elusive, as reported in a review by Casares \& Jonker (2014) where a deeply discussion about the mass measurements of SBHs and IMBHs is contained.

Supermassive black holes (SMBHs) are 10$^6$--10$^{10}$ times more massive than our Sun and are found in the centers of galaxies (see the exhaustive review by Kormendy \& Ho, 2013). Most galaxies, and maybe all of them, harbor such a black hole. So in our region of the Universe, there are some 10$^{11}$  SMBHs. The nearest one resides in the center of our Milky Way galaxy. The most distant one we know of resides in a quasar galaxy billions of lightyears away. SMBHs grow in size as they gorge on surrounding matter.
%%%%%%%%%%%%%%%%%%

A list of BH candidates has been reported by Robert Johnston (2004) and provides the input for constructing the map of sky locations of BH candidates. Source list includes results reported in Kormendy \& Gebhardt (2001), Orosz (2002), Tremaine et al. (2002), and Ziolkowski (2003).

%, as shown in Fig. 13. Stellar-mass black holes in red, intermediate-mass black holes in purple, supermassive black holes in blue. The base image is from Tycho sky map from JPL's Solar System Simulator.

%%%%%%%%%%%%%%%%%%% FIGURE 13 %%%%%%%%%%%%%%%%%
%\begin{figure*}%[!ht]%[!hbp]%[!ht]%[!hbp]%[h]
%\begin{center}
%\includegraphics[width=6cm]{BH_Candidates_Location.eps}
%\includegraphics[width=6.5cm]{COS-B_Catalog.eps}
%\includegraphics[width=7cm]{HEAO_A-1_Catalog.eps}
%\includegraphics[width=6.5cm]{3rd_EGRET_Catalog.eps}
%\caption{The map of sky locations of BH candidates. Red (stellar mass BHs), Purple (IMBHs),  Blue (SMBHs).
%The base image is from Tycho sky map from JPL's Solar System Simulator (adopted from Johnstone, 2004). }
%\end{center}
%\end{figure*}
%%%%%%%%%%%%%%%%%%%%%%%%%%%%%%%%%%%%%%%%%%%%%%

%%%%%%%%%%%%%%%%%%% FIGURE 13 %%%%%%%%%%%%%%%%%
\begin{figure*}%[!hbp]%[!ht]%[!hbp]%[!ht]%[!hbp]%[h]
\begin{center}
\includegraphics[width=12.5cm]{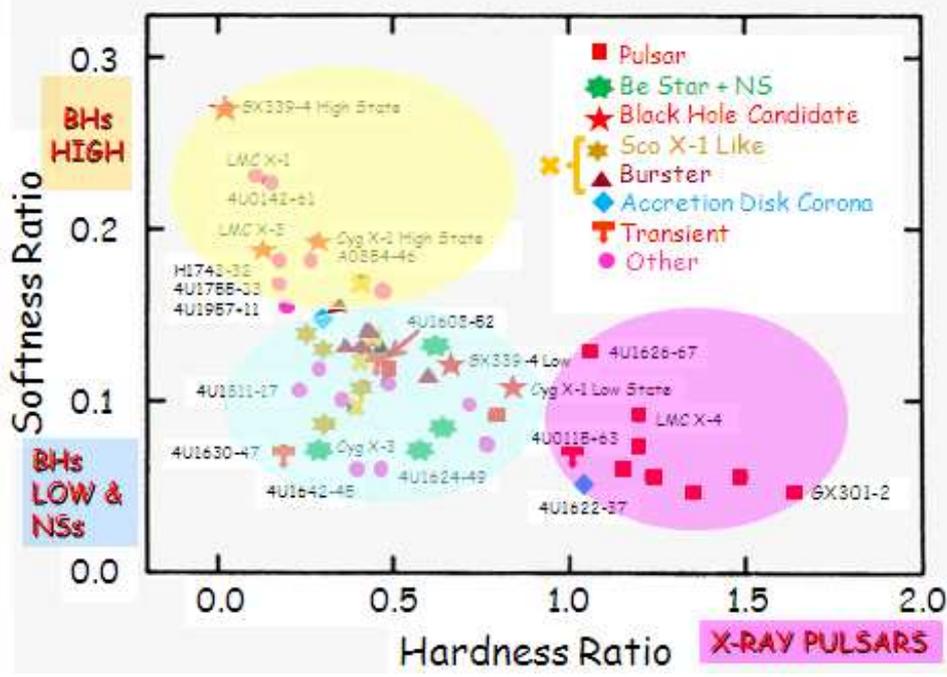}
\caption{Softness ratio versus hardness ratio for galactic compact systems. Light yellow ellipse marks the zone where BHs in high state lie, light turquoise ellipse marks the zone of the BHs in low state and NSs, and light fuchsia ellipse marks the zone of the X-ray pulsars (Giovannelli, 2016, after Yasuo Tanaka, 2001). }
\end{center}
\end{figure*}
%%%%%%%%%%%%%%%%%%%%%%%%%%%%%%%%%%%%%%%%%%%%%%

In the case of galactic compact sources, by using the softness and hardness ratios, coming for the measurements of the many X-ray satellites, it is possible to construct a diagram in which BHs in high state are separated by those in low state, and by other kind of objects, such as X-ray pulsars and other systems, as shown in Fig. 13 (Giovannelli, 2016, after Tanaka, 2001).

%\bigskip
%\noindent
%$^{(\ast)}$
%\footnotesize
%The Mach number is given by the ratio of the velocity of the gas to the local sound speed. In order to evaluate the temperature, concentration and velocity of the plasma near the black hole it is necessary to solve the system of equations describing the plasma motion (Michel, 1972) taking into account the distance from the black hole in units of gravitational radius and the u = R-component of four velocity. Mach's number 1 and 2 correspond to different values of u$_0^2$ (1.0266 and 2.1213, respectively) at a distance r$_0$ from the black hole (Giovannelli, Karaku{\l}a \& Tkaczyk, 1981, 1982).
%\normalsize
%\bigskip
%\noindent

\section{Model of time lag between optical and X-ray outbursts in AGNs}

%\subsection{Accretion onto black holes}
As already noted black hole accretion is a fundamental physical process in the universe. It is the
standard model for the central engine of active galactic nuclei (AGNs), and also plays a central role in the study of black hole X-ray binaries, Gamma-ray bursts, and tidal disruption events. According to the temperature of the accretion flow, the accretion models can be divided into two classes, namely cold and hot. The standard thin disk model belongs to the cold disk, since the temperature of the gas is far below the virial value (Shakura \& Sunyaev, 1973) (see reviews by (Pringle, 1981; Frank, King \& Rayne, 2002; Bisnovatyi-Kogan, 2011; Abramowicz \& Fragile, 2013; Blaes, 2014). The
disk is geometrically thin but optically thick and radiates multi-temperature black body spectrum.
The radiative efficiency is high, $\sim$ 0.1, independent of the accretion rate. The model has been successfully applied to luminous sources such as luminous AGNs and black hole X-ray binaries in the thermal state. The most recent review about the accretion onto black holes was published by Lasota (2016).

The tidal disruption of stars by massive BHs has been discussed since many years by Rees (1988), and e.g. Magorrian \& Tremaine (1999). Rees (1988) argued that stars in galactic nuclei can be captured or tidally disrupted by a central black hole. Some debris would be ejected at high speed, the remainder would be swallowed by the hole, causing a bright flare lasting at most a few years.
Such phenomena are compatible with the presence of 10$^6$--10$^8$ M$_\odot$ holes in the nuclei of many nearby galaxies.
Stellar disruption may have interesting consequences in our own Galactic Center if a $\approx 10^6$ M$_\odot$ hole lurks there.

In a recent paper, BKG17 developed models of time lags between optical and X-ray
flashes for close-binary galactic sources with accretion disks and for an AGN with an SMBH that is embedded in a quasi-spherical bulge. The flashes in an AGN are considered in the model when a disruption of a star that is in the evolution phase of a giant enters the radius of strong tidal forces. The matter with a low angular momentum that is released by the star falls into the SMBH in the form of a quasi-spherical flow with a velocity that is close to the
free-fall velocity. An X-ray flash occurs when the falling matter reaches the hot inner regions. The time lag observed in these sources is identified with the time of the matter falling from the tidal radius onto the central region. The values of the tidal radius that they calculated in this model were compared with the theoretical radii of a tidal disruption that depends on the masses of the SMBH and of the star, and on the radius of the star.

Knowing the SMBH masses from observations, and making a reasonable suggestion for the stellar mass that is on the order of one solar mass, they obtained that the radii of the disrupted star are between a few tens and a few hundreds of
R$_\odot$ (see Table 2). These radii are characteristic of stars of moderate mass on the giant phase of evolution (e.g. Bisnovatyi-Kogan, 2011).

The time delay between the optical and X-ray flashes is experimentally determined.
The radius at which the optical flash occurs is calculated for the motion with free-fall velocity V$_{\rm ff}$ as:

V$_{\rm ff}$  = (2GM/r)$^{1/2}$  ;  dr/dt = V$_{\rm ff}$  ; $\tau_{\rm ff}$ = 2/3 [r$^{3/2}$/(GM)$^{1/2}$]

\noindent
Taking $\tau_{\rm ff}$ = $\tau_{\rm obs}$ , BKG17 obtained a radius of the optical flash r$_{\rm opt}$ as:

r$_{\rm opt}$ = $1.65 \times 10^{12}$ $\tau_{\rm obs}$ m$^{1/3}$ cm \,\,\,\,\,\,\,\,\, (6.3)

\noindent
where: $\tau_{\rm obs}$ expressed in days, and SMBH mass: m expressed in (M$_\odot$).

%%%%%%%%%%%%%%%%%%%%% Table 2 %%%%%%%%%%%%%%%%%%%%%%%
%\footnotesize
\begin{table}[!hbp]
\begin{center}
\caption{Properties of stars tidally disrupted by SMBH in AGNs (adapted from Bisnovatyi-Kogan \& Giovannelli, 2017).
Here $m_s = \frac{\rm M_s}{\rm M_\odot}$, where ${\rm M_s}$ is the mass of the disrupted star.}
\bigskip
\begin{tabular}{|l|c|c|c|l|}
\hline
   &   &  & \\
Source name & $\tau_{\rm obs}$   & r$_{\rm opt}$ = r$_t$ & R$_s$ \\
        &   (days)     &  (cm)     &  (cm)   \\
\hline
Mrk 509 &  15  & $5.2 \times 10^{15}$ & $114 \times m_s^{1/3}$R$_\odot$\\
\hline
NGC 7469 & 4  & $8.95 \times 10^{14}$ & $47 \times m_s^{1/3}$R$_\odot$\\
\hline
3C 120 & 3.9-6.2?	(10)   & $3.3 \times 10^{15}$ & $100 \times m_s^{1/3}$R$_\odot$   \\
\hline
NGC 3516   &  100   & $1.1 \times 10^{16}$ & $409 \times m_s^{1/3}$R$_\odot$  \\
\hline
NGC 4051 & 2.4 & $2.5 \times 10^{14}$ & $36 \times m_s^{1/3}$R$_\odot$ \\
\hline
ASASSN-14li & 5 & $1.1 \times 10^{15}$ & $98 \times m_s^{1/3}$R$_\odot$ \\
\hline
\end{tabular}
\end{center}
\end{table}
%\end{footnotesize}

It is necessary to mention that another possibility to interpret the short time delays in AGNs is based on the irradiation model (e.g. Ulrich, Maraschi, \& Urry, 1997).
This model could explain recent extensive observations of NGC 5548 in X-rays (SWIFT), UV, and optical light (HST) in the "reverberation mapping" campaign (Edelson et al., 2015; Fausnaugh et al., 2016).

However, with the formula (6.3) BKG17 justify the experimental time delay between optical and X-ray flashes observed in AGNs.

\section{Discussion and Conclusions}

We have discussed the genesis of the work that allowed us to develop models of time lags between optical and X-ray
flashes for close-binary galactic sources with accretion disks and for an AGN with a SMBH that is embedded in a quasi-spherical bulge.

The time lag in disk-accreting galactic close-binary sources
is based on a sudden increase in the accretion flow that starts
at the disk periphery and is related to the optical maximum. The
massive accretion layer propagates to the central compact source
as a result of the turbulent viscosity. The X-ray flash occurs when
this massive layer reaches the inner hot regions of the accretion
disk and falls into the central compact object. The matter in the
accretion disk moves inside with a speed that is determined by
the turbulent viscosity. We described this model quantitatively
and derived an analytic formula that determines the value of the
time lag. This formula gives results that agree well with observational values.
The flashes in an AGN are considered in the model when a
disruption of a star that is in the evolution phase of a giant enters
the radius of strong tidal forces. The matter with a low angular
momentum that is released by the star falls into the SMBH in the
form of a quasi-spherical flow with a velocity that is close to the
free-fall velocity. An X-ray flash occurs when the falling matter reaches the hot inner regions. The time lag observed in these sources is identified with the time of the matter falling from the
tidal radius onto the central region. The values of the tidal radius
that we calculated in this model were compared with the theoretical radii of a tidal disruption that depends on the masses of the SMBH and of the star, and on the radius of the star. Knowing the SMBH masses from observations, and making a reasonable suggestion for the stellar mass that is on the order of one solar mass,
we obtained that the radii of the disrupted star are between a few tens and a few hundreds of R$_\odot$. These radii are characteristic of stars of moderate mass on the giant phase of evolution (see, for instance, Bisnovatyi-Kogan, 2011).

The matter with larger angular momentum that appeared in
the disruption of the star is expected to form an accretion disk
through which the matter will move to the center as a result of
turbulent viscosity, similarly to flashes in close galactic binaries.
This motion is much slower than free-fall velocity and may last
for many years. After such a flash in AGNs, we therefore expect
a long-duration irregular variability in the whole electromagnetic
spectrum.

The  variability  properties  observed  in  many  AGNs,  where
optical  and  UV  emission  lags  the  X-ray  light  curve,  may  be
explained  by  the  model  in  which  an  X-ray  flash  in  the  center
of AGN is followed by reradiation of the surrounding accretion
disk.

%\newpage
\bigskip
\bigskip
\noindent
{\bf Acknowledgments} This research has made use of the NASA's Astrophysics Data System;

\bigskip
\bigskip
\noindent {\bf DISCUSSION}

\bigskip
\noindent {\bf DMITRY BISIKALO:} What is the physical reason of mass-transfer changing in your "delay model"?

\bigskip
\noindent {\bf FRANCO GIOVANNELLI:} The physical reason for X-ray/Be is connected with eccentric orbit of NS or
BH in the binary system, where the accretion rate is increasing in the vicinity of the periastron of the
orbit. For CV the increase of the accretion rate is, probably, connected with the development of
instability in the outer parts of the accretion disk, leading to the turbulent state with high viscosity.

\bigskip
\noindent {\bf VICTOR DOROSHENKO:} You detected HeI line in 1999 in A0535+26, but there was no X-ray outburst, so there seems to be persistent disc around the NS, which is consistent with our XMM results in quiescence. What do you think about it?

\bigskip
\noindent {\bf FRANCO GIOVANNELLI:} Yes, it is possible. The INTEGRAL monitor also detected some activity in quiescence.

\bigskip
\noindent {\bf VICTOR DOROSHENKO:} Did I understand correctly that there is no need for field amplification for magnetars?

\bigskip
\noindent {\bf FRANCO GIOVANNELLI:} In our model the magnetic field is not important in the accretion disk
and accretion flow.
It determines only the inner boundary of the accretion disk around the NS.

\end{document}